\documentclass[preprint,12pt]{elsarticle}
\journal{Computer Physics Communications}
\usepackage{graphicx}
\usepackage{subcaption}
\usepackage{color}
\graphicspath{{figures/}{fig/}}
\usepackage{amsmath}
\usepackage{amssymb}
\usepackage{bm}
\usepackage{slashed}
\usepackage{epsfig}
\usepackage{amsfonts}
\usepackage{epstopdf}
\usepackage{hyperref}
\usepackage{bbm}
\usepackage{textcomp}
\usepackage{color}
\usepackage{tabularx,makecell,multirow}
\usepackage{algorithm}
\usepackage{algpseudocode}
\usepackage{capt-of}
\usepackage{caption}
\usepackage{bigints}

\newcommand{\mi}{\mathrm{i}}
\newcommand{\md}{\mathrm{d}}
\newcommand{\ep}{\epsilon}
\newcommand{\bl}{{\tt Blade}}

\newcommand*\circled[1]{\raisebox{.5pt}{\textcircled{\raisebox{-.9pt} {#1}}}}

\begin{document}
\begin{frontmatter}

\title{{\tt Blade}: A package for block-triangular form improved Feynman integrals decomposition}

\author[a,b]{Xin Guan}
\ead{guanxin0507@pku.edu.cn}

\author[c]{Xiao Liu}
\ead{xiao.liu@physics.ox.ac.uk}

\author[a,d]{Yan-Qing Ma}
\ead{yqma@pku.edu.cn}

\author[a,e]{Wen-Hao Wu}
\ead{wuwenhao21@mails.ucas.ac.cn}


\address[a]{School of Physics, Peking University, Beijing 100871, China}
\address[b]{SLAC National Accelerator Laboratory, Stanford University, Stanford, CA 94039, USA}
\address[c]{Rudolf Peierls Centre for Theoretical Physics, Clarendon Laboratory, Parks Road, Oxford OX1 3PU, UK}
\address[d]{Center for High Energy Physics, Peking University, Beijing 100871, China}
\address[e]{School of Astronomy and Space Science, University of Chinese Academy of Sciences (UCAS), Beijing, China}

\begin{abstract}
	In this article, we present the package {\tt Blade} as the first implementation of the block-triangular form improved Feynman integral reduction method. The block-triangular form has orders of magnitude fewer equations compared to the plain integration-by-parts system, allowing for strictly block-by-block solutions. This results in faster evaluations and reduced resource consumption.
	We elucidate the algorithms involved in obtaining the block-triangular form along with their implementations. Additionally, we introduce novel algorithms for finding the canonical form and symmetry relations of Feynman integrals, as well as for performing spanning-sector reduction. Our benchmarks for various state-of-the-art problems demonstrate that {\tt Blade} is remarkably competitive among existing reduction tools.
	Furthermore, the {\tt Blade} package offers several distinctive features, including support for complex kinematic variables or masses, user-defined Feynman prescriptions for each propagator, and general integrands.
\\

\noindent \textbf{PROGRAM SUMMARY}

\begin{small}
\noindent
{\em Program title:} {\tt Blade}\\
{\em Developer's repository link:} \url{https://gitee.com/multiloop-pku/blade}\\
{\em Licensing provisions:} MIT\\
{\em Programming language:} {\tt Wolfram Mathematica} 11.3 or higher\\
{\em External routines/libraries used:} {\tt Wolfram Mathematica} [1], {\tt FiniteFlow} [2]\\
{\em Nature of problem:}
Automatically reducing dimensionally regularized Feynman integrals into linear combination of master integrals.\\
{\em Solution method:}
The program implements recently proposed block-triangular form to significantly improve the reduction efficiency.\\
{\em Restrictions:} the CPU time and the available RAM \\
{\em References:}
{\\} [1] \url{http://www.wolfram.com/mathematica}, commercial algebraic software;
{\\} [2] \url{https://github.com/peraro/finiteflow}, open source.
\end{small}

\end{abstract}


\end{frontmatter}

\newpage

\vspace{1cm}

\newpage

\tableofcontents

\section{Introduction}
Precision physics has been playing an indispensable role in the interpretation of particle phenomena and in identifying deviations from the Standard Model predictions.
As the experimental precision will significantly increase due to the good performance of Large Hadron Collider \cite{Azzi:2019yne,Cepeda:2019klc} and the proposed next generation colliders \cite{ILC:2013jhg,Behnke:2013xla,Bambade:2019fyw,CEPCStudyGroup:2018ghi,CEPCStudyGroup:2018rmc,TLEPDesignStudyWorkingGroup:2013myl,FCC:2018byv,FCC:2018evy}, theorists have to calculate high order perturbative corrections of scattering amplitudes to make precise theoretical predictions so as to exploit the full potential of experiments. The computation of Feynman integrals (FIs) presents one of the obstacles to this end.
The most widely adopted strategy for computing scattering amplitudes is to reduce FIs in amplitudes to some so-called master integrals by integration-by-parts (IBP) reduction method \cite{Chetyrkin:1981qh,Tkachov:1981wb} and then evaluate master integrals.
Both of the two steps are very challenging for state-of-the-art calculations. It is important to note that the reduction is also a key ingredient in many methods to calculate master integrals.

IBP reduction combined with Laporta algorithm \cite{Laporta:2000dsw} is now almost ubiquitously used and there exist many public packages \cite{Anastasiou:2004vj,
	Peraro:2019svx,
	Smirnov:2008iw,Smirnov:2013dia,Smirnov:2014hma,Smirnov:2019qkx, Lee:2013mka, Smirnov:2023yhb, Wu:2023upw, Maierhofer:2017gsa, Maierhofer:2018gpa,Klappert:2020nbg,Studerus:2009ye,vonManteuffel:2012np}.
The main bottleneck of the IBP reduction is the memory usage and time-consumption due to huge size of linear equations.
In the past few years, we have witnessed many improvements to overcome these difficulties.
The finite field technique avoids the \textit{intermediate expression swell} and reduces the memory consumption significantly \cite{vonManteuffel:2014ixa}, which has been widely used in frontier calculations and implemented in several packages \cite{Peraro:2016wsq,  Peraro:2019svx, Klappert:2019emp, Klappert:2020aqs,  Belitsky:2023qho}.
The syzygy equations \cite{Gluza:2010ws,Schabinger:2011dz,Cabarcas:2011ddd,Larsen:2015ped,Ita:2015tya,Zhang:2016kfo,Abreu:2018rcw,Bohm:2018bdy,vonManteuffel:2020vjv,Badger:2021imn,Abreu:2021asb} can trim IBP identities, ensuring, for example, that propagators with dots do not occur, consequently significantly reducing the size of IBP system.
It was found that a good choice of master integrals, such as UT basis \cite{Henn:2013pwa,Abreu:2018rcw,Abreu:2018zmy,Abreu:2021asb} and D-factorized basis \cite{Smirnov:2020quc, Usovitsch:2020jrk}, can greatly simplify the final results of reduction, thus accelerate the reduction itself.
It has been observed that rational functions in multi-loop calculations can be significantly simplified via partial fraction \cite{Boehm:2020ijp,Bendle:2021ueg,Heller:2021qkz} and a method of reconstructing rational functions in partial-fractioned form was proposed recently \cite{Chawdhry:2023yyx}.

We note that IBP system is very sparse and involves many auxiliary Feynman integrals.
An efficient reduction can be achieved if a well-structured and compact linear system can be found.
The method of block-triangular form \cite{Liu:2018dmc,Guan:2019bcx}, developed by three of the authors, is along this line.
The block-triangular form usually has orders of magnitude fewer equations than traditional IBP system and they can be solved strictly block by block, resulting in faster numeric evaluation and reduced resources-consumption.
It has already been applied to several state-of-the-art calculations \cite{Chen:2022mre,Chen:2022vzo,Chen:2023osm,Guan:2023gsz,Bi:2023bnq}.

In this article, we present \bl, the first public implementation for the \textbf{BL}ock-tri\textbf{A}ngular form improved Feynman integral \textbf{DE}composition.
\bl~typically enhances IBP reduction efficiency by 1-2 orders, thereby rendering many intricate physical problems achievable now.
Furthermore, the \bl~package offers several other distinctive features. Firstly,  it accommodates both real and imaginary components in kinematic variables or masses. Secondly,  it enables us to differentiate propagators with different Feynman prescriptions, allowing us to introduce their respective symmetries individually. Thirdly, it supports IBP reduction of integrals with general integrands as long as the derivatives of the integrand with respect to loop momenta can be expressed as linear combinations of such integrands, with propagators as coefficients. 

The rest of the paper is organized as follows. In Section \ref{sec:ibp-reduction-method} we review the IBP reduction method and introduce its implementations in \bl.
In Section \ref{sec:block-triangular-form}, we delve into detailed discussion of the block-triangular form method and its implementations in \bl.
The benchmarks are presented in Section \ref{sec:benchmarks}.
In Section \ref{sec:summary}, a summary is given.

\section{Integration-By-Parts reduction}\label{sec:ibp-reduction-method}

\subsection{Feynman integrals}

A {scalar FI} with $L$ loop momenta $l_i~(i=1,\cdots,L)$ and $E$ independent external momenta $p_i~(i=1,\cdots,E)$ is usually defined as
\begin{equation}\label{eq:def_integrand}
	I(\mathbf{\nu}) = \int \left( \prod_{i=1}^L \frac{\md^d l_i}{\mi \pi^{d/2}} \right)  \frac{D_{k+1}^{-\nu_{k+1}} \ldots D_{N}^{-\nu_N}}{D_1^{\nu_1} \ldots D_k^{\nu_k}} ,
\end{equation}
where $d$ represents the space-time dimension, $N={L(L+1)}/{2}+LE$ denotes the number of loop-momentum-dependent scalar products ($l_i\cdot l_j$ and $l_i \cdot p_j$), $D_a$ for ${a=1,\cdots,k}$ represents propagator denominators, and $D_a$ for ${a=k+1,\cdots, N}$ is referred to as irreducible scalar product (ISP). In this context,  $D_a$ ($a=1,\cdots,N$) are linear combinations of $l_i\cdot l_j$ and $l_i \cdot p_j$, and the introduction of ISPs facilitates the expression of $l_i\cdot l_j$ and $l_i \cdot p_j$ as linear combinations of $D_a$.
FIs encompassing all feasible values of $\nu_a$,  adhering to the condition that ${\nu}_i \in \mathbb{Z}$ for $i=1, \cdots, N$ and $ \nu_i < 0$ for $i=k+1,\cdots,N$, are referred to as an \textit{integral family} essentially defined by the propagator denominators.

An integral family can be categorized into different sectors. A \textit{sector} comprises FIs $I(\mathbf{\nu})$ that share the same vector
\begin{align}
	\Theta(\mathbf{\nu}) = \{ \theta(\nu_1),\ldots,\theta(\nu_N) \},
\end{align}
where
\begin{align}
	\theta(\nu_i) = \left\{ \begin{matrix} 1, &\enskip \nu_i > 0 \\ 0, &\enskip \nu_i \leq 0 \end{matrix} \right. \,.
\end{align}
In essence, integrals within the same sector exhibit identical set of denominators.
For two sectors $A$ and $B$, if $\Theta(\mathbf{\nu}^{(A)})\neq\Theta(\mathbf{\nu}^{(B)})$ and $\theta(\nu_a^{(A)}) \leq \theta(\nu_a^{(B)})$ for all $a$, then sector $A$ is termed a \textit{sub-sector} of $B$ and sector $B$ is termed a \textit{super-sector} of $A$. The set of denominators of a sub-sector is a subset of those in the original sector, illustrating that diagrammatically, a sub-sector can be obtained by contracting propagators of the Feynman diagram of the original sector.
As there is always a maximal sector in an integral family with respect to this order, this sector is denoted as the \textit{top-sector} of the family. Conversely, an \textit{integral family} comprises of a top-sector and all its sub-sectors.

In \bl, we introduce more general integrand, which  multiplies the integrand in Eq.~\eqref{eq:def_integrand} by another non-polynomial scalar factor  $\mathcal{F}_i$,
\begin{align}\label{eq:generalintegrand}
	\frac{\mathcal{F}_i}{\prod_{a=1}^{N} D_a^{\nu_a}}, \;\, \text{for } i=1, \cdots, n,
\end{align}
where  $\mathcal{F}_i$ satisfies the condition that $\frac{\partial \mathcal{F}_i}{\partial D_a}$ can be re-expressed as linear combination of terms in the form of Eq.~\eqref{eq:generalintegrand} with coefficients independent of loop momenta.
This general integrand has several applications, such as symbolic reduction and generating functions for FIs \cite{Guan:2023avw,Feng:2022hyg}. Even though introducing an additional factor, we still use the same terminologies sector and family for integrals with integrand in Eq.~\eqref{eq:generalintegrand}.

As integrals, FIs are unchanged under reparametrization, or transformation, of integration variables $l$ to $l'=\phi(l_1,\cdots,l_L,p_1,\cdots,p_E)$ where $\phi$ can be any function. Although complete freedom exists in transformation, our focus will be solely on scenarios that establish connections between distinct FIs within a specific family. This implies that the integrand, following a transformation and multiplication by the Jacobian, can be reformulated as a linear combination of the integrand defined in Eq.~\eqref{eq:generalintegrand}.
The transformations that meet this criterion can be divided into two distinct categories: infinitesimal transformations and finite transformations. We will discuss these two categories separately in the following two sections.

\subsection{Infinitesimal transformations: IBPs, LIs, and zero sectors}

Infinitesimal transformations have the general form
\begin{align} \label{eq:inftrans}
	l_i^\mu\to  l_i^\mu+ \lambda \phi_i^\mu(l),
\end{align}
where $\lambda$ is an infinitesimal constant. Under the infinitesimal transformation,  the loop integration measure $\md\mu_{L} \equiv \prod_{i=1}^L \frac{\md^d l_i}{\mi\pi^{d/2}}$ transforms as
\begin{align}
	\md\mu_{L} \to \md\mu_{L} \left( 1+ \lambda \sum_{i=1}^L\frac{\partial \phi_i^\mu(l)}{\partial l_i^\mu} \right) ,
\end{align}
and a given  integrand $f(l)$ transforms as
\begin{align}
	f(l) \to f(l)+ \lambda  \sum_{i=1}^L \phi_i^{\mu}(l) \frac{\partial f(l)}{\partial l_i^\mu} .
\end{align}
Note that $\mathcal{O}(\lambda^2)$ terms have been neglected here and in the following.
As the integral is unchanged under the transformation, we have
\begin{align}
	&\int \mathrm{d}\mu_{L} f(l)\nonumber\\
	=& \int \md\mu_{L} \left( 1+ \lambda \sum_{i=1}^L \frac{\partial \phi_i^\mu(l)}{\partial l_i^\mu} \right) \left( f(l)+ \lambda \sum_{i=1}^L \phi_i^{\mu}(l) \frac{\partial f(l)}{\partial l_i^\mu} \right)\nonumber\\
	=&\int \md\mu_{L}  \left( f(l)+ \lambda \sum_{i=1}^L \frac{\partial }{\partial l_i^\mu } \bigg( f(l) \phi_i^{\mu}(l)\bigg) \right),
\end{align}
which results in
\begin{align}\label{eq:IBPs}
	&\int \md\mu_{L} \sum_{i=1}^L   \frac{\partial }{\partial l_i^\mu } \bigg( f(l) \phi_i^{\mu}(l)\bigg)=0.
\end{align}
By selecting appropriate $\phi_i$ so that the integrand in the above equation can be represented as a linear combination of the terms defined in Eq.~\eqref{eq:generalintegrand},  a linear relationship between FIs can be established. The above equation is widely recognized as the IBP identity~\cite{Chetyrkin:1981qh,Tkachov:1981wb}, as it can be derived through the application of integration by parts to reformulate the left-hand side in terms of surface terms, which ultimately vanish in the context of dimensional regularization.

Our derivation demonstrates that all relations between FIs resulting from infinitesimal transformations can be constructed solely using IBP relations. As a result, this kind of relations constructed from other ways seeming different from IBPs are redundant. Even though, it was found that some other relations are useful to improve the efficiency of IBP reductions in practice, including the Lorentz invariant identities (LIs) and relations defining zero sectors.

Infinitesimal Lorentz transformation  for external momentum $p_i$ can be expressed as $p_i^{\mu} \to p_i^{\mu}+ \lambda \omega^{\mu\nu} p_{i\,\nu}$, where $\omega^{\mu\nu}$ is antisymmetric between $\mu$ and $\nu$. Because scalar FIs are unchanged in Lorentz transformation, we have
\begin{align}
	&\int \mathrm{d}\mu_{L} f(l)= \int \md\mu_{L}  \left( f(l)+ \lambda  \omega^{\mu\nu} \sum_{i=1}^E p_{i\,\nu} \frac{\partial f(l)}{\partial p_i^\mu} \right).
\end{align}
Considering the antisymmetric nature of $\omega^{\mu\nu}$, the coefficients of independent components of $\omega^{\mu\nu}$ give rise to the well-known LIs \cite{Gehrmann:1999as}
\begin{align} \label{eq:LIs}
	p_j^\mu p_k^\nu \sum_{i=1}^{E} \left(p_{i\nu} \frac{\partial}{\partial p_{i}^\mu} - p_{i\mu} \frac{\partial}{\partial p_{i}^\nu} \right)\int \mathrm{d}\mu_{L}  f(l)=0,
\end{align}
where a factor $p_j^\mu p_k^\nu$ is multiplied to obtain scalar relations.
LIs can be derived from IBPs because, at the integrand level, the  Lorentz transformation for external momenta $p_i^{\mu} \to p_i^{\mu}+ \lambda \omega^{\mu\nu} p_{i\,\nu}$ is equivalent to perform the Lorentz transformation for loop momenta $l_i^{\mu} \to l_i^{\mu} - \lambda \omega^{\mu\nu} l_{i\,\nu}$ with external momenta untouched. By substituting $\phi_i^\mu(l)=-\omega^{\mu\nu} l_{i\,\nu}$ into Eq.~\eqref{eq:IBPs}, the coefficients of independent components of $\omega^{\mu\nu}$ give rise to
\begin{align} \label{eq:IBP2}
	p_j^\mu p_k^\nu\int \mathrm{d}\mu_{L} \sum_{i=1}^{L} \left( \frac{\partial}{\partial l_{i}^\mu} l_{i\nu} - \frac{\partial}{\partial l_{i}^\nu}  l_{i\mu} \right)f(l)=0,
\end{align}
which is equivalent to Eq.~\eqref{eq:LIs} based on the above argument, consistent with the observation made in Ref.~\cite{Lee:2008tj}.
However, while they are equivalent, Eq.~\eqref{eq:IBP2} incorporates more intricate FIs that ultimately cancel out during the summation process, rendering it less practical for use.

If all FIs in a given sector are evaluated to zero using IBP relations, that sector is referred to as a \textit{zero sector}. Identifying and eliminating these zero sectors in advance can significantly improve the efficiency of IBP reductions. The presence of zero sectors is a direct consequence of the scaleless nature of some FIs, which means that under a specific reparametrization, all FIs in the given sector remain unchanged apart from acquiring an overall non-unit factor. Here we will only consider FIs which do not have  non-polynomial scalar factor  $\mathcal{F}_i$ in the integrand, otherwise they are usually not scaleless. Since FIs with ISPs can be represented as linear combinations of FIs without ISPs but with a shifted spacetime dimension \cite{Davydychev:1991va,Tarasov:1996br,Tarasov:1997kx}, it is adequate to prove that all FIs lacking ISPs vanish in a particular sector to demonstrate that it is a zero sector. To this end, a straightforward algorithm is presented in \cite{Lee:2013mka}, utilizing the Lee-Pomeransky representation for FIs:
\begin{equation}
	I(\nu) = \frac{\Gamma\left(\frac{d}{2}\right)}{\Gamma\left(\frac{(L+1)d}{2}-\sum_{i=1}^K\nu_i\right)} \bigintsss \left(\prod_{i=1}^K \frac{\mathrm{d} z_i z_i^{\nu_i-1}}{\Gamma(\nu_i)}\right) G(z)^{-d/2},
\end{equation}
where $G(z)=F(z)+U(z)$ represents the summation of Symanzik polynomials \cite{Lee:2013hzt}, and we have assumed that the sector under consideration has $K$ propagator denominators.

Now considering the rescaling transformation
$ z_i \to z_i^\prime= z_i+\lambda\, c_iz_i$, where $c_i$ are constants and $\lambda$ is a freely chosen parameter,
the criterion for zero sector is that $G(z)$ also gains an overall factor:
\begin{align} \label{eq:Gz}
	G(z) \to G(z^\prime)= h(\lambda)\, G(z),
\end{align}
where $h(\lambda)$ is independent of $z$ and, by normalizing $c_i$, we can alway assume $h(\lambda)=1+\lambda+\mathcal{O}(\lambda^2)$ for small $\lambda$.
Indeed, if the criterion is satisfied,the rescaling transformation results in the relation
\begin{align}
	I(\nu)=\prod_{i=1}^N (1+\lambda\, c_i)^{\nu_i-1} h(\lambda)^{-d/2} I(\nu),
\end{align}
which necessitates that $I(\nu)=0$ for any value of $\nu$.
The criterion Eq.~\eqref{eq:Gz} is equivalent to its infinitesimal form, where the nontrivial relation is
\begin{equation}
	\sum_{i} c_i z_i \frac{\partial G(z)}{\partial z_i} = G(z).
\end{equation}
This equation can be recast as a set of linear equations involving the unknown constants $c_i$. Solving these equations, if a solution exists, indicates that the sector is indeed a zero sector.

\subsection{Finite transformation: canonical form and symmetry relations}\label{sec:finitetransf}

For finite transformations, our focus is solely on the general linear transformation given by
\begin{equation}\label{eq:loop-trans}
	l_i \to l_i' = \sum_{j=1}^{L} A_{ij} l_j + \sum_{j=1}^E B_{ij} p_j,
\end{equation}
where matrices $A$ and $B$ are independent of loop momenta. The reason for this restriction is that nonlinear transformations would alter the quadratic structure of the inverse propagator denominators in loop momenta, rendering them no longer quadratic, and thus will not generate useful relations. By substituting the linear transformation given in Eq.~\eqref{eq:loop-trans} into FIs without non-polynomial scalar factor  $\mathcal{F}_i$, we obtain the following relation:
\begin{equation}
	\begin{aligned}
		&\int \left( \prod_{i=1}^L \frac{\mathrm{d}^d l_i'}{i \pi^{d/2}} \right) \left( \prod_{a=1}^{N} D_a(l')^{-\nu_a} \right) \\
		&= \left| \det A \right| \int \left( \prod_{i=1}^L \frac{\mathrm{d}^d l_i}{i \pi^{d/2}} \right) \left( \prod_{a=1}^{N} D'_a(l)^{-\nu_a} \right).
	\end{aligned}
\end{equation}
Here, $D_a(l')$ represents the $a$-th propagator denominator expressed in terms of the transformed loop momenta $l'$, while $D'_a(l)$ is the corresponding denominator in the original loop momenta $l$ after applying the transformation. The absolute value of the determinant of the matrix $A$, denoted as $\left| \det A \right|$, arises from the change of variables in the integration.
This relation formally connects two FIs that are defined differently in terms of their loop momenta. Repeatedly performing such transformations allows us to navigate through a space of equivalent representations for FIs, effectively mapping them to different ``orbits" within this space.
Within this framework, we can identify a particular representative point for each orbit that serves as a standard or canonical form for the corresponding equivalent class of FIs. By mapping all FIs to their respective canonical forms, we can accurately identify equivalent FIs, referred to as symmetry relations relating FIs in different sectors.

We define the canonical form based on the denominators $D_a$ with positive indices $\nu_a > 0$. Specifically, for each loop momentum $l_i$ ($i=1, \cdots, L$), we require that either $l_i^2-m^2$ or a linear term $l_i\cdot q$ (where $m$ represents the mass associated with the propagator denominator and $q$ is a linear combination of external momenta) appears as one of these denominators as far as possible. These $L$ constraints limit the freedom of linear transformations, resulting in a finite number of candidate representations that satisfy these criteria. The subsequent step involves arranging these candidates in a predefined order to ultimately select the final representation. However, enumerating and sorting all candidates is impractical due to their vast quantity. Fortunately, the transformations relating these representations constitute a well-organized discrete group. By harnessing the structure of this group, we can effectively find out the representative one without enumerating all of them.

Sometimes, there exist transformations that leave the canonical form unchanged, which means that the set of denominators with positive indices are mapped into the same set. Mathematically speaking, these transformations constitute an isotropy subgroup within the larger transformation group. Since these transformations map FIs to the same sector, the isotropy group gives rise to symmetry relations among FIs within that particular sector. For more details, refer to Appendix \ref{app}, where we introduce an algorithm tailored to efficiently determine the canonical form of any given FI, provide transformations that relate the FI to its canonical form, and thus generate all possible symmetry relations.\footnote{We note that an alternative way to find symmetry relations is to use the Pak's algorithm \cite{Pak:2011xt,Gerlach:2022qnc}, based on the parameterized representations.}

Unconstrained phase space integrations can be conceptualized as loop integrations, provided that a delta function is included as part of the propagators to impose the on-shell condition for the phase space momentum. However, with fixed on-shell conditions, phase space momenta cannot undergo linear transformations in the same manner as loop momenta described in Eq.~\eqref{eq:loop-trans}. Instead, when we permute the phase space momenta which have the same invariant mass, it does not alter the final outcome. This characteristic gives rise to additional symmetry relations.
In our approach, we enumerate all possible permutation scenarios for phase space momenta. For each scenario, we determine its corresponding pre-canonical form by applying the linear transformation of loop momenta in Eq.~\eqref{eq:loop-trans}. Subsequently, we arrange these pre-canonical forms in a predefined order and select the last one as our final canonical form.

Occasionally, permutations of certain unintegrated external momenta leave all scalar products unchanged. For instance, when there are only two independent, unintegrated, light-like external momenta, swapping them preserves both their light-like nature and their scalar product. In such cases, the value of the FIs remains unaffected by these permutations.
Analogous to the symmetries arising from phase space momenta, we enumerate all possible permutation scenarios for the unintegrated external momenta. For each scenario, we determine its corresponding pre-canonical form by combining the linear transformations of loop momenta with all feasible permutations of phase space momenta. We then arrange these pre-canonical forms according to a predefined order and select the last one as our final canonical form.
This structured approach enables to identify all potential symmetry relations. 

We also identify additional symmetries in which the denominators of Feynman integrals (FIs) depend on a subset of kinematic variables. Moreover, permutations of certain unintegrated external momenta leave these relevant kinematic variables unchanged. Graphically, this corresponds to a factorized diagram with some loop momenta decoupled or an intrinsic lower-point diagram after contracting propagators, where some unintegrated external momenta are grouped. For each factorized component, we redefine each combination of grouped external momenta as relevant external momenta.
Next, we enumerate all possible permutations for the relevant external momenta and determine symmetries as discussed in the previous paragraph. Extra symmetries have a one-to-one correspondence among denominator propagators (positive indices). However, for numerator propagators (negative indices), we cannot express them as linear combinations of the complete set of propagators of the integral family because the entire set of kinematic variables changes under permutations of relevant external momenta. Instead, we perform tensor decomposition to separate out all dependencies on non-relevant external momenta from loop integrals \footnote{A different approach is used in Ref. \cite{Wu:2024paw}.}. Then, we apply permutations of relevant external momenta solely to loop integrals and reorganize the obtained FIs.

\subsection{Laporta algorithm}

For a long time, the ultimate goal of FIs reduction has been to solve IBP, LI and symmetry identities algebraically to find a complete set of reduction rules, known as \emph{recurrence relations}, with symbolic powers $\nu$. However, the construction procedures are obscure and success is not guaranteed.
The Laporta algorithm  \cite{Laporta:2000dsw} provides a systematic reduction strategy  with numerical $\nu$. Within this algorithm,
one choices some specific integer values for
$\nu$, referred to as \textit{seeds}, to set up a system of equations. Solving this system leads to a reduction of target integrals to master integrals.

\subsubsection{Integral ordering}
The integral ordering defines which FIs are complicated and which ones are simple. The goal is to reduce complicated FIs to simple ones, which are defined as master integrals, as much as possible.
The complexity of the Gaussian elimination algorithm for the IBP system can also strongly depend on the integral ordering.
A good choice of order can retain the sparsity of the system during the Gaussian elimination, resulting in faster evaluation.

To classify and sort FIs, it is customary to associate each FI with the following numbers:
\begin{itemize}
	\item $t$ is the total number of positive indexes, 
	\begin{equation}
		t = \sum_{j|\nu_j>0} 1\,.
	\end{equation}
	\item $d$ is the number of \textit{dots} of a FI, defined as
	\begin{equation}
		d = \sum_{j|\nu_j>0} (\nu_j-1)\,.
	\end{equation}
	\item $r$ is the \textit{rank} of a FI, defined as the opposite value of the summation of the negative indexes,
	\begin{equation}
		r =- \sum_{j|\nu_j<0} \nu_j\, .
	\end{equation}
\end{itemize}

There are four predefined integral orderings in \bl,
\begin{enumerate}
	\item $t \succ d \succ r$,(default)
	\item $t \succ r \succ d$,
	\item $t \succ (d+r) \succ d \succ r $,
	\item $t \succ (d+r) \succ r \succ d$.
\end{enumerate}
where $\succ$ means the importance of the criteria, in descending ordering. For all of the four orderings, FIs in sub-sectors are considered simpler than that in the top sector. The ordering 1 prefers to define bases with no increased propagator powers, by requiring that, for instance, FIs in the same sector are considered to be more complicated if they have more dots.
When employing the ordering 1, if master integrals with increased propagator powers emerge, it typically suggests that the Feynman integrals are not fully reduced yet.
Users can modify the integral ordering via the option ``IntegralOrdering'' within the function ``BLSetReducerOptions''.

\subsubsection{Observations on seeds}\label{sec:seeds}
We refer to the representative sector within an equivalent class of sectors related by symmetry relations as a unique sector, and those that can be mapped to the unique sector as mapped sectors. By default, the IBP system generated by \bl~consists of IBP, LI and symmetry identities for seed integrals within unique sectors; symmetry identities relating different sectors for seed integrals in mapped sectors; and extra symmetry identities for maximal cut master integrals.

There is no strategy to determine priorly a minimal set of seed integrals that can reduce a given set of target integrals to master integrals.
As mentioned in \cite{Peraro:2019svx}, it is often convenient to specify a maximal value of $r$ and $d$, and, in most cases, one only needs seed integrals with $d$ either the same or one unity larger than the maximal one among the target integrals.
This non-minimal strategy is adopted in \bl.

In physical applications, one usually needs to reduce a large number of FIs with either high rank or many dots.
For instance, after applying the reverse unitarity technique \cite{Anastasiou:2002yz} for real emission processes, the FIs belonging to top sector exhibit a higher rank without dot while FIs in sub-sectors possess a lower rank but a few dots.
The IBP system becomes excessively large if we distribute seeds with equal rank and dots across each sector, determined by the maximal rank and dots of the target integrals.
In \bl, we partition the target integrals into a few sub-families using the following algorithm:
\begin{enumerate}
	\item Select the most complicated FI among the target FIs and use it to define a sub-family. Then generate a set of FIs in this sub-family, labeled by $G$, by operating on target FIs that share the same sector and dots as the most complicated FI, with the property that FIs having smaller values of $t$ also having smaller values of $r$.\label{enum:dividelevel-pick} 
	\item Exclude $G$ from target integrals and proceed to step \ref{enum:dividelevel-pick} until all the target integrals have been allocated into respective sub-families.
	\item Trim the IBP systems of each sub-family. 
	\item Finally, either conduct reduction within sub-families separately and merge results afterward, or merge  the trimmed IBP systems of all sub-families and then trim it again and solve it.  \label{enum:dividelevel-merge}
\end{enumerate}
By employing this algorithm, we group integrals according to distinct features, such as high rank and high dots, leading to a more efficient seeding strategy. 

Moreover, with the property of FIs in $G$, the rank of seeds in lower-level sectors can be chosen to be lower than those in higher-level sectors.
This is accomplished by introducing the option ``FilterLevel'' for the function ``BLSetReducerOptions'' to adjust the rank of seeds.\footnote{Similar idea was also described in \cite{Driesse:2024xad}}
Given an initial rank $r$ and assuming the top-sector to have $t$ propagators, setting ``FilterLevel'' as $\{c_0, \ldots,c_k\}$ will modify the rank of seed integrals as follows:
\begin{itemize}
	\item For integrals with $t$ propagators, the rank will be adjusted to $r+c_0$,
	\item For integrals with $t-1$ propagators, the rank will be adjusted to $r+c_1$,
	\item ...,
	\item For integrals with $t-k$ or fewer propagators, the rank will be adjusted to $r+c_k$.
\end{itemize}
The default value of ``FilterLevel'' is $\{0,0,0,-1\}$. (Alternatively written as 3, interpreted as a list with the first 3 elements as zeros and the 4th element as -1). This strategy significantly reduces the IBP system’s size and memory usage since lower-level sectors play a major role in compiling IBP identities.

Notice that we can be more bold, for instance by setting ``FilterLevel'' to $\{1, 0, -1, -2\}$.
Assigning fewer seeds to lower-level sectors further reduces the total number of equations dramatically. Giving more seeds to higher-level sectors helps reduce the most complicated integrals, while maintaining the total number of equations almost unchanged. This strategy may sometimes yield very good results. 
We recommend that users first test reducing target integrals with moderate ranks to gain experience before attempting high-rank reductions.

There is a competition regarding whether to generate IBP and LI identities for mapped sectors. While generating IBP and LI can yield easier equations, resulting in faster numerical evaluation after trimming, it also increases the size of the system before trimming, particularly when mapped sectors constitute a significant portion of the topology. This effect may prevail, leading to increased memory usage, rendering trimming impractical. By default, \bl ~doesn't generate IBP and LI identities for mapped sectors, but we recommend to generate these identities by setting ``IBPForMapSector'' to {\it True} via the function ``BLSetReducerOptions'', provided that there is enough memory. Furthermore, symmetry relations for seed integrals of very high rank can be too complicated to parse and solve, leading to increased  memory usage and time-consumption. Hence, it is worthwhile to assess the necessity of generating symmetry relations, particularly for seed integrals of very high rank. Users can close symmetry by setting ``CloseSymmetry'' to {\it True} via the function ``BLSetReducerOptions''.

\subsubsection{Determine master integrals}\label{sec:determine-master-integral}
It has been proved that the number of master integrals is finite for each given family \cite{Smirnov:2010hn}. The next question is determining the number of master integrals, which helps affirm whether the reduction is complete and estimate the complexity of integral families, aiding in assessing the feasibility of tackling a physical process.
It was pointed out in Ref. \cite{Lee:2013hzt} that the number of master integrals is determined from the critical points of the polynomials entering either the parametric representation or equivalently the Baikov representation\cite{Baikov:1996iu} of the integral, and the method has been implemented in a {\tt Mathematica} package {\tt Mint}.
In Ref.~\cite{Bitoun:2017nre}, it was pointed out that the number of master integrals can be computed by the Euler characteristic of the Lee-Pomeransky polynomial.

Alternatively, the number of mater integrals as well as a possible choice of mater integrals can be determined using the maximal cut technique, which means that, for each sector, we perform IBP reduction by ignoring all FIs in subsectors. This method is efficient because the IBP system is much smaller for each maximal cut. This idea has been implemented in the {\tt Singular/Mathematica} package {\tt Azurite} \cite{Georgoudis:2016wff}. In \bl, similar idea has been employed.
Our algorithm can be summarized as follows:
\begin{enumerate}
	\item Initialize the starting power to $p$ and set the integral ordering locally to either the default ordering 3 or 4.
	\item Select integrals that meet the condition $r+d < p$ to serve as both seed integrals and target integrals. Generate IBP and LI identities on maximal cut with respect to seed integrals and perform the reduction. \label{enum:maximalcut-preduce}
	\item 
    Increase $p$ and repeat step \ref{enum:maximalcut-preduce} until reaching a level of Feynman integrals where all sharing the same value of $(r+d)$ can be fully reduced. The master integrals are denoted as $M_1$. \label{enum:maximalcut-auto}
	\item Generate IBP and LI identities to reduce $M_1$ to $M_2$, such that $M_2$ adheres to the user-defined integral ordering.\label{enum:maximalcut-basistrans}
	\item Apply symmetry relations to further reduce $M_2$ to a possible smaller set $M_3$.\label{enum:maximalcut-symmetry}
\end{enumerate}
The criterion of complete reduction outlined in step \ref{enum:maximalcut-auto} is based on empirical observations and has proven universally applicable across encountered cases.
It is plausible because the criterion is sufficient to construct a closed differential equation with respect to  all internal particle masses. This approach offers the advantage of eliminating the necessity for manual setup of seed and target integrals.
The step \ref{enum:maximalcut-basistrans} and \ref{enum:maximalcut-symmetry} aim to obtain a maximal cut basis whose structure is similar to the master integrals in the full IBP reduction.
For instance,  if user-defined integral ordering is the default ordering 1, $M_2$ has no increased propagators powers, in contrast to $M_1$ where in general consist of FIs with both nonzero rank and dots. $M_3$ will be finally added into the set of user-defined target integrals, and, after solving the system of IBP relations, it will check whether the resultant master integrals can be expressed as linear combinations of $M_3$.
If this criterion is not met, it indicates insufficient seeds, prompting \bl ~to automatically increase the number of seeds.

The maximal cut method facilitates an automatic seeding program and, since its time consumption is small compared to full reduction, it is activated by default. To deactivate it, utilize the function ``BLSetReducerOptions'' and set ``CheckMastersQ'' to {\it False}.

\subsection{Spanning-sector reduction}
The \textit{Generalized-cut} method\cite{Larsen:2015ped, Britto:2024mna} is widely applied to identify coefficients of master integrals.  An advantage of applying cuts is that any Feynman integral lacking a cut propagator can be set to zero under the cut. Thus, the IBP system for each cut can be much smaller than the complete one. 
However, it is common that, at the integral level, IBP can produce magic relations that relates a sector A to other sectors B, where B are neither equal to A nor sub-sectors of A. 
Magic relations emerge only in the presence of super sectors, becoming apparent after Gaussian elimination of IBP identities associated with those super sectors (see \cite{Smirnov:2013dia,Maierhofer:2018gpa} for further discussions). Magic relations spoil the structure where integrals are reduced to their sub-sector integrals, leading to issues with the generalized cut reduction method.
Alternatively, it is well known that one can reduce the memory usage of IBP reduction by setting some master integrals to zero, either all except one (``master-wise") or all except those in the same sector (``sector-wise")\cite{Chawdhry:2018awn, Klappert:2020nbg}.
However, intuitively, as we compute the reduction coefficients of a sub-sector master integral, we may have already computed the coefficients of its super sectors in the intermediate step. So, it would be beneficial to set a suitable chosen set of sectors to zero and avoid duplicate computations. 

Inspired by generalized-cut and master-wise method, we propose a spanning-sector reduction algorithm as follows:
\begin{enumerate}
	\item Use the maximal-cut technique to determine master integrals, denoted as $M_{raw}$. Subsequently, extend $M_{raw}$ to obtain a set of sample integrals, denoted as $G$. Numerically reduce $G$ to master integrals, referred to as $M_{nosym}$, ensuring that $M_{nosym}$ remains a subset of $M_{raw}$. Symmetry is closed in this step. \label{item:mnosym}
	
	\item Identify the lowest-level sector in $M_{nosym}$ and gather both the sector itself and its super-sectors, denoted as $T$. \label{enum:spanning0}
	
	\item From the reduction results in step \ref{item:mnosym}, identify FIs in $G$ that are outside sectors $T$ but have non-vanishing reduction coefficients on master integrals in $T$. Denote these as $G_{mag}$.
    Sectors related to magic relations, denoted as $T_{mag}$, are sectors defined by FIs in $G_{mag}$. If $T_{mag}$ is empty, then $T$ forms a spanning-sector, meaning FIs in other sectors can be set to zero if we are only interested in reduction coefficients for master integrals in $T$. Proceed to step \ref{enum:exclude}. If $T_{mag}$ is not empty, incorporate $T_{mag}$ and its super-sectors into $T$ and proceed to step \ref{enum:spanning}. 	\label{enum:spanning}
	
	\item Exclude FIs in $T$ from $M_{nosym}$ and proceed to step \ref{enum:spanning0} until $M_{nosym}$ is empty, at which point a minimal set of spanning-sectors is obtained. \label{enum:exclude}
	
	\item Perform integral reduction within each spanning-sector (without symmetry) and merge the reduction tables to get the complete reductions.\label{enum:reduce_span}
	
	\item Apply symmetry relations to reduce master integrals in the step \ref{enum:reduce_span} to a possible smaller set $M_{sym}$, thus expressing the complete reduction results as linear combinations of $M_{sym}$.
\end{enumerate}
In step \ref{item:mnosym}, the set of sample integrals $G$ consists of FIs from each sector with ranks that are one higher than the maximum rank of the maximal cut masters in that sector. For reducible sectors, sample integrals have ranks up to 1. If $M_{nosym}$ is a proper subset of $M_{raw}$, this indicates the presence of magic relations.
In the step \ref{enum:spanning}, the spanning-sector is equivalent to a generalized cut if there is no magic relations. In the step \ref{enum:reduce_span}, we set those sectors not belonging to the spanning-sector to zeros in the first place and don't generate IBP identities for them, resulting in significant memory reduction. Note that we assume---but have not proved---that all magic relations are fully identified in the first step, and no new magic relations will be generated in the step \ref{enum:reduce_span}. Otherwise, for some spanning-sectors, the number of master integrals obtained in the step \ref{enum:reduce_span} will be smaller than that in the $M_{nosym}$. In such cases, \bl ~would report an error and return \$Failed. To activate the spanning-sector reduction algorithm, set the option ``SpanningReduce'' to {\it True} in the function ``BLReduce''.

\subsection{Finite field technique}\label{sec:finite-field}

Finite-field methods have been widely adopted in FI reduction \cite{vonManteuffel:2014ixa,Peraro:2016wsq}. In this algorithm, reductions are performed numerically\footnote{Here and throughout the rest of the paper, unless otherwise specified, ``numerical'' refers to rational numbers over a finite field of a large prime number.} in a finite field and the analytic form of the final result is constructed with interpolation techniques. Subsequently, we repeat the process for a few primes until the exact rational numbers can be obtained via the Chinese remainder theorem. This approach avoids the cumbersome manipulation of large rational numbers during intermediate steps in direct rational computation.

Another important application of finite-field is to \textit{trim} the equations generated from the seeds. By solving the system numerically, one can determine a smaller set of independent equations which is sufficient to reduce target integrals.
\bl ~uses {\tt FiniteFlow} \cite{Peraro:2019svx} as a sparse solver and finite-field reconstructor.

\subsection{Syzygy equations}
If we substitute the integrand from Eq.~\ref{eq:def_integrand} into Eq.~\ref{eq:IBPs}, it becomes apparent that IBP identities may involve integrals with dots greater than seed integrals by 1, stemming from the derivative of denominators with respect to loop momenta. However, many of these integrals do not contribute to amplitudes. Hence, it would be beneficial to choose $\phi_i^\mu$ so that  IBP identities do not increase the number of dots, by requiring the following syzygy equations \cite{Gluza:2010ws},
\begin{align}
	\sum_i \phi_i^\mu \frac{\partial D_j}{\partial l_i} = \gamma_j D_j,
\end{align}
where both $\phi_i^\mu$ and $\gamma_j$ are polynomials of propagators. Syzygy equations can be solved either through linear algebra by making ansatzes for $\phi_i^\mu$ and $\gamma_j$\cite{ Cabarcas:2011ddd, Schabinger:2011dz} or through algebraic geometry\cite{Larsen:2015ped,Ita:2015tya, Bohm:2017qme,Bohm:2018bdy}. The optimised system of IBP relations through the solution of syzygy equations makes their solution substantially simpler.

\section{Block-triangular form}\label{sec:block-triangular-form}
IBP reduction is usually one of the main bottlenecks in frontier problems.
One main reason for the inefficiency of IBP reduction is that IBP system involves huge number of linear equations, but most of them are used to solve irrelevant auxiliary FIs.
The key idea of the block-triangular form is to construct linear relations within a much smaller chosen set of integrals, and the relations are chosen in the way that they can be strictly solved block by block.

The block-triangular form realizes a step-by-step reduction in the sense that the most complex integrals are reduced to simpler integrals in the first block and these simple integrals are further reduced to even simpler integrals in other blocks.
Eventually, all integrals can be reduced to master integrals, which, by definition, are the simplest FIs.

\subsection{Search algorithm}

\subsubsection{Two-step search strategy}

A two-step search strategy to construct relations among Feynman integrals was developed in Refs. \cite{Liu:2018dmc,Guan:2019bcx}.

~\newline
\noindent \textit{Step 1}

In the first step, we set up a system of relations that can numerically express all target integrals in terms of master integrals:
\begin{equation}\label{eq:vector-rep}
	I_i(\ep,\vec{s})=\sum_{j=1}^n C_{ij}(\ep,\vec{s}) M_j(\ep, \vec{s}),
\end{equation}
where $\epsilon = \frac{4-d}{2}$ is the dimensional regulator and $\vec{s}$ represents the kinematic invariants in the problem. The system is allowed to be somewhat inefficient in numerical calculations; thus, the system is not required to be block-triangular. This system can be obtained either by using the $\eta$ series representation of Feynman integrals \cite{Liu:2018dmc}, or simply by using the IBP method. In the context of \bl, we use the numerical IBP method which has been  well-studied. 

~\newline
\noindent \textit{Step 2}

In general, a linear relation among Feynman integrals $G \equiv \{I_1,I_2,\dots,I_N\} $ can be expressed as:
\begin{equation}\label{eq:general-rel}
	\sum_{i=1}^N Q_i(\ep,\vec{s})I_i(\ep,\vec{s})=0,
\end{equation}
where the coefficients $Q_i(\ep,\vec{s})$ can be decomposed into linear combinations of polynomials with respect to $\ep$ and $\vec{s}$:
\begin{equation}\label{eq:ansatz-of-rel}
	Q_i(\ep,\vec{s}) = \sum_{\mu_0=0}^{\ep_{max}}\sum_{\vec{\mu}\in \Omega_{d_i}} \tilde{Q}_i^{\mu_0\mu_1\dots\mu_r} \ep^{\mu_0} s_1^{\mu_1} \dots s_r^{\mu_r},
\end{equation}
where $\tilde{Q}_i^{\mu_0\mu_1\dots\mu_r}$ are rational unknowns to be determined, $\ep_{max}$ is an integer that specifies the maximal degree of $\ep$, $\Omega_{d_i} = \{\vec{\mu}\in \mathbb{N}^r |\mu_1+\cdots+\mu_r=d_i\}$, $d_i$ is the half of the mass dimension of $Q_i(\ep,\vec{s})$, and the maximal degree of $\vec{s}$ is determined by $d_{max}=\max(d_i)$.

After substituting Eq.~\eqref{eq:vector-rep} and Eq.~\eqref{eq:ansatz-of-rel} into Eq.~\eqref{eq:general-rel}, we obtain
\begin{equation}
	\sum_{\mu_0,\vec{\mu}} \sum_{j=1}^n \tilde{Q}_i^{\mu_0 \mu_1 \dots \mu_r}
	\ep^{\mu_0} s_1^{\mu_1} \dots s_r^{\mu_r} C_{ij}(\ep,\vec{s}) M_j(\ep,\vec{s})= 0.
\end{equation}
The coefficient of each MI in the above equation must vanish because the MIs are linearly independent. This leads to $n$ constraints among $\tilde{Q}_i^{\mu_0 \mu_1 \dots \mu_r}$,
\begin{equation}\label{eq:search_constraints}
	\sum_{\mu_0,\vec{\mu}}
	\tilde{Q}_i^{\mu_0 \mu_1 \dots \mu_r}
	\ep^{\mu_0} s_1^{\mu_1} \dots s_r^{\mu_r} C_{ij}(\ep,\vec{s}) = 0,~ j=1,\dots,n.
\end{equation}
where all quantities except $\tilde{Q}_i^{\mu_0 \mu_1 \dots \mu_r}$ are numerical.
It is important to note that for any given $\ep_{max}$ and $d_{max}$, the number of $\tilde{Q}_i^{\mu_0 \mu_1 \dots \mu_r}$ is finite. By repeating the aforementioned procedure at different numeric points, we can obtain adequate constraints to determine the unknowns $\tilde{Q}_i^{\mu_0 \mu_1 \dots \mu_r}$.
Since the above calculation is carried out within the finite field of a large prime, a rational reconstruction applying the Chinese remainder theorem is necessary to achieve the block-triangular form within the rational field.
Therefore, all linear relations among $G$ for any given $\ep_{max}$ and $d_{max}$ can be fully determined.

We split the set of integrals into $G\equiv G_1 \bigcup G_2$, where $G_1$ consists of FIs that are more complex than that in $G_2$.\footnote{The definition of complexity is a consequence of a convention to order integrals.}
To reduce $G_1$ to $G_2$, we increase the degree bound ($\ep_{max}$ and $d_{max}$) and search relations among $G$ until there are enough relations to express $G_1$ in terms of $G_2$. These relations form a block.

\subsubsection{Polynomial ansatz}
As in the Eq.~\eqref{eq:ansatz-of-rel}, one needs to make polynomial ansatz for the expected relations among Feynman integrals. We can always set one variable with a non-vanishing mass dimension to 1 and recover its dependency at the end. Hence, we assume the problem depends on $n$ dimensionless variables $z_1,z_2,\dots,z_n$.
To accommodate general scenarios, \bl ~allows users to freely formulate polynomial ansatzes. This is facilitated by utilizing four variables: ``VariableGroup", ``VariableWeight", ``IntegralWeight" and ``CutIncrement", which can be configured via the function ``BLSetSearchOptions''.

The term ``VariableGroup" refers to a list of variable groups. For instance, variables can be divided into two groups $\{\{z_1,\dots,z_k\},\{z_{k+1},\dots,z_n\}\}$. This partition enables users to set up different polynomial ansatz modes for different variable groups, while their direct products form the final polynomial ansatz, as depicted in Eq.~\eqref{eq:ansatz-of-rel}.
Eq.~\eqref{eq:ansatz-of-rel} is a special case where the dimensional regulator $\ep$ and the kinematic invariants $\vec{s}$ are divided into two groups.

Next, we focus on the polynomial ansatz mode for a specific group of variables, such as $\{z_1,\dots,z_k\}$. ``VariableWeight" is a list of integers denoted as $\vec{\omega}$, whose length corresponds to the number of variables. For a given degree bound $\tilde{d}$, the possible polynomial ansatz can be written as
\begin{equation} \label{eq:degree-bound-tilde}
	\{z_1^{\mu_1} \cdots z_k^{\mu_k}\}|_{\vec{\mu}\in \mathbb{N}^k \bigwedge \sum_{j=1}^k \mu_j \omega_j \leq \tilde{d}}
\end{equation}
For instance, if $\vec{\omega}=\{1,2\}$ and $\tilde{d}=3$, all possible values of $\vec{\mu}$ are
\begin{equation}
	\{\{0,0\},\{1,0\},\{0,1\},\{2,0\},\{1,1\},\{3,0\}\}
\end{equation}
It is evident that the lower the variable weight, the greater the potential complexity within the polynomial ansatz.
Thus, the ``VariableWeight" can be configured based on the variable's complexity.

The crucial question then arises: how do we assess the complexity of variables? This can be achieved by means of the \textit{univariate block-triangular form}. Using numerical IBP as input while preserving $z_2=z_{2,0},\dots,z_n=z_{n,0}$, we can construct the univariate block-triangular form. Here, the coefficients of Feynman integrals manifest as linear combinations of $z_1$ monomials.
We extract the maximal degree of $z_1$, denoted as $m_1$, from the univariate block-triangular relations. $m_1$ is referred to as the lower bound of $z_1$ because the maximal degree of $z_1$ within the multivariate-block-triangular form must be greater than or equal to $m_1$.
Using the method of the univariate block-triangular form, we obtain the lower bounds of $z_i$ variables, denoted as $\vec{m}$. A higher lower bound for a variable denotes a higher level of complexity.
It is worth noting that obtaining the univariate block-triangular form is relatively easy because the number of unknowns in Eq.~\eqref{eq:ansatz-of-rel} is small and the intricate dependency of other variables simplifies to a rational number.

There are two modes in \bl ~for determining ``VariableWeight", called \textit{uniform} and \textit{adaptive},
\begin{equation}
	\omega_i =
	\begin{cases}
		1 & \textit{uniform}, \\
		\begin{cases}
			\left. \text{floor}(\frac{\max(\vec{m})}{m_i})\right., & \text{if} ~{m_i\neq 0}  \\
			\left. 1000 \max(\vec{m})\right., &\text{if} ~{m_i=0}
		\end{cases}
		& \textit{adaptive}.
	\end{cases}
\end{equation}
where $\vec{m}$ is the lower bounds of $z_i$ variables, obtained by univariate block-triangular form. Users can specify exact integer for ``VariableWeight" as well.

We introduce ``IntegralWeight" to specify the weight, denoted as $\omega(I_i)$, of a FI contributing to the degree of a linear relation among FIs.
Given a relation degree $d$, the possible polynomial ansatzes for the coefficients of Feynman integral $I_i$ are
\begin{equation}
	\left. \{z_1^{\mu_1}\cdots z_k^{\mu_k}\}\right|_{\vec{\mu}\in \mathbb{N}^k \bigwedge \sum_{j=1}^k \mu_j \omega_j +\omega(I_i)\leq d}
\end{equation}\label{eq:possible_ansatz}
Note that $d$ differs from $\tilde{d}$ that serves as the maximal degree of the polynomial ansatz in Eq.~\eqref{eq:degree-bound-tilde}.
There are two modes in \bl ~to determine ``IntegralWeight", defined as \textit{uniform} and \textit{dimension},
\begin{equation}
	\omega(I_i) =
	\begin{cases}
		0, & \textit{uniform}, \\
		\mathrm{dim}[I_i] -\min(\mathrm{dim}[I_j]|_{I_j \in G}), & \textit{dimension}.
	\end{cases}
\end{equation}
where $``\mathrm{dim}"$ denotes the half of the mass dimension, and $G$ encompasses all integrals in the relation.
If ``VariableWeight" is set to ``uniform" and ``IntegralWeight" is set to ``dimension", we can get the polynomial ansatz concerning $\vec{s}$ in Eq.~\eqref{eq:ansatz-of-rel}. 

We need to set up ``VariableWeight" and ``IntegralWeight" for each variable group.
Then the program need to automatically increase the degree bound until the obtained relations facilitate  reduction. Increasing degree bound is straightforward when dealing with only one group of ``VariableGroup".
For multiple variable groups, \bl ~adopts a strategy where it sets the degree bounds to all except the last one variable group.
Given ``VariableGroup" as $\{\vec{x}_1,\dots,\vec{x}_r\}$ and degree bounds as $\{c_1, c_2, \ldots, c_{r-1}\}$, Blade begins by incrementally increasing the degree of $\vec{x}_1$ until it reaches its bound $c_1$. Then, it increments the degree of $\vec{x}_2$ by one unit and resets the degree of $\vec{x}_1$ from 0 to $c_1$. This process continues, with each group's degree being incremented until it reaches its respective bound. Once a group reaches its bound, the procedure moves to the next variable group and repeats the process.
For example, if there are two variable groups and the degree bound is $\{3\}$, then the incrementation scheme would be as follows:
\begin{align}
	\{&\{0,0\},\{1,0\},\{2,0\},\{3,0\}, \nonumber \\
	& \{0,1\},\{1,1\},\{2,1\},\{3,1\},\nonumber \\
	& \{0,2\},\{1,2\},\cdots\nonumber\\
	& \qquad \quad \quad \quad ~~\cdots \quad \quad \quad \quad\}\, .
\end{align}

It is crucial to choose proper degree bounds such that \bl ~efficiently explore the solution space in a multi-direction manner.
To this end, we introduce the concept of \textit{single-group block-triangular form} and ``CutIncrement". The single-group block-triangular form is analogous to univariate block-triangular form, with the distinction that one group of variables remains analytic.
By employing the single-group block-triangular form method, we can ascertain the lower bound of each variable group, denoted as $\{m_1,\dots,m_r\}$.
Each degree bound must exceed or equal to the corresponding lower bound.
Note that the single-group block-triangular form, such as respecting $\vec{x}_1$, may correspond to relations with intricate dependency on other variable groups, such as $\vec{x}_2,\ldots,\vec{x}_r$. This intricate dependency is simplified to a single number in this case, facilitating successful search.
Therefore, when other variable groups are restored to being analytic, it is advantageous to slightly raise the degree bound based on the lower bound. This adjustment aims to search for simpler relations overall, with considerably simplified dependencies on $\vec{x}_2,\ldots,\vec{x}_r$ and a slightly higher degree on $\vec{x}_1$.
We introduce ``CutIncrement" as $\{i_1,\dots,i_{r-1}\}$, and the degree bounds are calculated as $\{m_1+i_1,\dots,m_{r-1}+i_{r-1}\}$.

For broader applicability, we configure the default search options as
\begin{align*}
	``\mathrm{VariableGroup}" & \rightarrow \{\{z_1,\cdots,z_n\}\},\\
	``\mathrm{VariableWeight}"  & \rightarrow \{adaptive\},  \\
	``\mathrm{IntegralWeight}"
	&\rightarrow \{uniform\}, \\
	``\mathrm{CutIncrement}" &\rightarrow \{2,\cdots,2\}.
\end{align*}

It is notable that if there exists a relation such as
\begin{align}
	z_1 z_2 I_1 + z_2^2 I_2 + \cdots =0.
\end{align}
any monomials of $\vec{z}$ multiplied by this  relation would generate new relations with higher degrees. However, these new relations are redundant, and they need to be removed beforehand.
Using this strategy can significantly reduce the number of polynomial ansatzes thus enhancing search efficiency.
When proceeding to a different prime field or  a fixed phase space point, we leverage the knowledge gained from the search phase, wherein we retain only non-vanishing polynomial ansatz that yield independent relations among Feynman integrals(the block-triangular form).
Fewer polynomial ansatzes lead to an enhanced speed of solving the constraints (Eq.~\eqref{eq:search_constraints}) as well as a reduced number of numerical IBP.
We refer to the subsequent construction of the block-triangular form as \textit{fitting}.

\subsubsection{Algorithm}

We summarize our search algorithm as follows.

\begin{center}
	\renewcommand{\figurename}{Algorithm}
	\captionof{figure}{Search algorithm}\label{alg:search_algorithm}
	\fbox{
		\begin{minipage}{\dimexpr\columnwidth-2\fboxsep-2\fboxrule}
			\begin{algorithmic}[1]
				\Require Sort masters integrals in the order of descending complexity. The number of master integrals is denoted as $m$, The number of numerical IBP probes is denoted as $p$
				\Ensure Upper-right triangular form
				\For{$i \gets 1$ \textbf{to} $m$}
				\State  flag=True
				\State $j \gets 1$
				\While{$j < p$ \textbf{and} flag}
				\State Generate the constraint from the $i$-th master integral and the $j$-th numerical IBP probe, denoted as $E_{ij}$
				\For{$k \gets 1$ \textbf{to} $i-1$}
				\For{$l \gets 1$ \textbf{to} $n_k$}
				\State Substitute $E_{kl} \rightarrow E_{ij}$
				\EndFor
				\EndFor
				\For{$l \gets 1$ \textbf{to} $j-1$}
				\State Substitute $E_{il} \rightarrow E_{ij}$
				\EndFor
				\If{$E_{ij}$ is trivial}
				\State flag=False
				\Else
				\State Normalize $E_{ij}$
				\State $j$ ++
				\EndIf
				\EndWhile
				\State $n_i \gets j-1$
				\For{$j \gets n_i $ \textbf{to} 1}
				\For{$l \gets n_i$ \textbf{to} $j+1$}
				\State Substitute $E_{il} \rightarrow E_{ij}$
				\EndFor
				\EndFor
				\EndFor
			\end{algorithmic}
		\end{minipage}
	}
\end{center}

It is notable that by sorting the master integrals, the closer the master integral is positioned to the front, the fewer integrals have a non-zero projection onto it. This implies that the equations generated earlier, as depicted in line 5, are sparser.
From line 6 to line 13, we utilize all prior equations to simplify the newly generated equation, known as forward elimination.
By this procedure, the sparsity of the system is retained to a large degree.
In line 14, an equation is called `trivial' if it equals zero, indicating that the corresponding master integral can no longer provide an independent constraint, despite an increase in numerical IBP probes.
Conversely, if the equation isn't trivial, we normalize it in line 17 such that its leading coefficient equals 1, and then proceed to the next numerical IBP probe in line 18.
Furthermore, from line 22 to line 26, we conduct a backward elimination within equations generated from a specific master integral, resulting in a reduced row-echelon form. This reduces the complexity of forward elimination to $\mathcal{O}(c N^2)$, where $N$ represents the number of equations and $c$ signifies the average number of entries in each equation after backward elimination---approximately the dimension of solution space.
The complexity of backward elimination can be estimated at around $\sim m b^2 (b + c) \sim N^3/m^2 + c N^2/m $, where $b\sim N/m$ stands for the average number of equations generating from one single master integral. As a result, the total computational complexity amounts to roughly  $\sim c N^2 + N^3/m^2$. Despite the algorithm's $N^3$ component, its coefficient is comparatively small, resulting in a dominant $N^2$ algorithm. For example, in a typical configuration where $c\sim 100$, $N\sim 10^5$, $m \sim 100$, the ratio between the two components is 10.

The nullspace of the obtained upper-right triangular system can be easily determined, hence the relations among FIs. These relations would be validated by another numerical IBP probe.

\subsection{Reduction scheme}\label{chap:integral-extension}
Reduction scheme determines the set of integrals which will be involved in each block.
It is crucial to choose a proper set of FIs such that the resulting block-triangular form is simple and easy to search.
To this end, \bl ~adopts two methods, denoted as \textit{operator extension} and \textit{global extension}, to operate on target integrals and return a set of integrals among which the block-triangular form are constructed.

We introduce the notation of \textit{generalized-sector} for FIs:
\begin{align}
	\Theta^G(\mathbf{\nu}) = \{ \theta^G(\nu_1),\ldots,\theta^G(\nu_N) \},
\end{align}
where
\begin{align}
	\theta^G(\nu_i) = \left\{ \begin{matrix} \nu_i, &\enskip \nu_i > 0 \\ 0, &\enskip \nu_i \leq 0 \end{matrix} \right. \,.
\end{align}
For two generalized sectors $A$ and $B$, if $\Theta^G(\mathbf{\nu}^{(A)})\neq\Theta^G(\mathbf{\nu}^{(B)})$ and $\theta^G(\nu_a^{(A)}) \leq \theta^G(\nu_a^{(B)})$ for all $a$, then $A$ is termed a \textit{generalized sub-sector} of $B$ and $B$ is termed a \textit{generalized super-sector} of $A$.

The \textit{operator extension} is a generalization of $\hat{m}^\circleddash$ as defined in Ref.~\cite{Guan:2019bcx}. When acting on a target integral associated with a generalized sector $\Theta^G(\mathbf{\nu})$, having $t_0$ positive indexes and a rank of $r_0$, the operator extension yields all integrals belonging to $\Theta^G(\mathbf{\nu})$ and its generalized sub-sectors, with the condition that the rank of an integral with $t$ positive indexes should be not greater than $(r_0 - t_0 + t)$.

The \textit{global extension} yields all integrals belonging to $\Theta^G(\mathbf{\nu})$ and its generalized sub-sectors, with each integral possessing a rank no greater than $m$, where $m$ is an optional parameter, configured using the option ``MinimalSchemeRank'' within the function ``BLSetSchemeOptions'' (1 by default). The global extension complements the operator extension, particularly when the rank of the target integrals is low. In such cases, the extended integral set provided by the operator extension may be too small to form simple block-triangular relations, necessitating the inclusion of more integrals.

It is customary to construct blocks based on sectors. Within each sector, we denote target integrals in this sectors (excluding master integrals) as $G_1$. We apply both the operator extension and the global extension to $G_1$, to obtain an integral set $G$. Using the search algorithm described before, we search relations among $G$ to find out sufficient independent relations that can express $G_1$ as linear combinations of other integrals.

Note that symmetry relations and magic relations can spoil the structure where integrals are reduced to their sub-sector integrals, indicating that the relations among FIs generated by applying the operator extension and the global extension on $G_1$ are inadequate for reducing $G_1$. This prompts \bl ~to include more integrals into the block.

In practice,  \bl ~begins by applying the integral extension on target FIs, yielding a set of FIs denoted by $S$. Subsequently, $S$ is reduced to master integrals through IBP reduction. When constructing blocks, for each sector, those integrals not belonging to master integrals serve as target integrals, denoted as $G_1$. \bl ~identifies master integrals on which $G_1$ depends by solving IBP system numerically. The master integrals, which share the same number of propagators as that of $G_1$, are denoted as $M$. The integral set $G$ for reducing $G_1$ comprises all FIs in $S$ satisfying the condition that they belong to either the generalized (sub-)sectors of $G_1$ 
or generalized (sub-)sectors of $M$. 

We can discard mapped sub-sectors in $G$ to speed up the search with the option ``UniqueSubsectorQ''$\rightarrow$ True. To maintain the numerical efficiency of solving the block-triangular form, we need to control the block size. Integrals within one sector can be distributed across several blocks based on distinct generalized sectors with the option ``PartitionDot'' $\rightarrow$ True. Additionally, we can divide a single block into two blocks: one for reducing integrals with the highest rank and the other for reducing the remaining integrals. This partition becomes active when the block size reaches the optional threshold ``PartitionRankThreshold''. These three options can be configured via the function ``BLSetSchemeOptions''.

\bl ~also supports reducing linear combinations of FIs. We associate to each combination a ``sector'' based on the sector of its most complex component FI. Combinations sharing the same sector are reduced within the same block. The integral set of the block-triangular form is obtained by applying the previously mentioned integral extension methods to each component FI. We note that there is a way to further optimize the integral set for reducing derivatives of master integrals. We apply the integral extension to component FIs with a rank no greater than the maximum rank of the master integrals, resulting in fewer integrals in the block and speeding up the search.

\subsection{Adaptive search strategy}
Recall that the multivariate polynomial reconstruction can be accomplished through recursive application of Newton's formula. Indeed, a multivariate polynomial in $z_1,\ldots,z_n$ can be seen as a function of $z_1$ with coefficients being functions of $z_2,\ldots,z_n$,
\begin{align}
	f(z_1,\ldots,z_n)=\sum_{r=0}^{R} a_r(z_2,\ldots,z_n) \prod_{i=0}^{r-1}(z_1 - y_i).
\end{align}
For a fixed value of $z_2,\ldots,z_n$, one can perform univariate reconstruction to determine coefficients $a_r$. This means that the problem of $n$-variate polynomial reconstruction is transformed into the problem of $(n-1)$-variate polynomial reconstruction. Applying the strategy recursively, one arrives at the univariate polynomial reconstruction of $z_n$. Here comes the key observation: the numerical points needed for multivariate polynomial reconstruction form a hierarchical structure. For each $z_1$, there exist numerous ${z_2, \ldots, z_n}$, and within these, identical $z_2$ leads to numerous ${z_3,\ldots, z_n}$. The hierarchical structure also applies to multivariate rational function reconstruction (at least in {\tt FiniteFlow} \cite{Peraro:2019svx}). Additionally, it is worth noting that a list of sample points can be obtained based on the degree information learned from univariate rational reconstruction. This enables us to make full use of computing resources and evaluate sample points independently before starting any multivariate reconstruction.

Based on the structure of sample points, alternatively, we can search for the block-triangular form with a fixed value of $z_1$, using a small number of numerical IBP points as input ($z_1$ fixed, $z_2,\ldots,z_n$ varying), and then using the much more efficient block-triangular form to compute a large number of samplings ($z_1$ fixed, $z_2,\ldots,z_n$ varying). By searching for and solving a series of block-triangular forms under distinct values of $z_1$, we can finally compute all sample points and reconstruct the function. Actually, two extreme cases involve fixing $\{\}$ and fixing $\{z_1,\ldots,z_n\}$, corresponding to full-analytic block-triangular form and numerical IBP, respectively. We refer to the block-triangular form with fixed values of some variables as the `semi-analytic' block-triangular form.

The question is: which block-triangular form is better? Given that the total number of samplings is constant and the numerical efficiency of the semi-analytic block-triangular forms is comparable, there is a balance between the total number of numeric IBP (as input) and the time required to search for the block-triangular form. As we introduce another analytic variable $z_r$ into the block-triangular form, the search becomes more time-consuming due to the larger space of polynomial ansatzes regarding $z_r$. The number of numerical IBP for a single block-triangular form also increase, approximately proportional to the increase in polynomial ansatzes. However, the total number of numerical IBP decreases significantly due to the reduced total number of block-triangular forms. The reduction in the total number of block-triangular forms typically has a greater effect than the increase in numerical IBP in a single block-triangular form, because the reduction in the former is proportionate to the total degree of $z_r$, while the total degree of $z_r$ is generally much greater than the increase in space of polynomial ansatzes.

To determine the optimal strategy, we propose the adaptive search algorithm as outlined below:

\begin{center}
	\renewcommand{\figurename}{Algorithm}
	\captionof{figure}{Adaptive search}\label{alg:adaptive_search}
	\fbox{
		\begin{minipage}{\dimexpr\columnwidth-2\fboxsep-2\fboxrule}
			\begin{algorithmic}[1]
				\Require The estimated real time $T_{0}$ for computing probes using numeric IBP in one finite field, time factor $F$, a list of search variables $z_1,\ldots,z_n$
				\Ensure The optimal level of block-triangular form
				\State $T \gets T_{0}$
				\State $i \gets 0$
				\Repeat
				\State $i \gets i + 1$
				\State Search the block-triangular form with analytic coefficients in $z_{n-i+1},\dots,z_n$, while holding $z_1,\dots,z_{n-i}$ fixed at a numeric value. The search phase should not surpass $F \times T$ in real time; if it does, abort when it reaches that limit. Record the time-consumption as $t_{i}$.
				\If{$t_{i} = F \times T$}
				\State $T_{i} \gets \infty$
				\Else
				\State Estimate the real time, denoted as $T_{i}$, for computing probes using the block-triangular form at this level (analytic $z_{n-i+1},\dots,z_n$).
				\EndIf
				\State $T \gets \min(T_{i},T)$
				\Until{$T_{i} = \infty$ or $i = n$}
				\State \textbf{return} the value of $j$ where $T_{j} = T$.
			\end{algorithmic}
		\end{minipage}
	}
\end{center}

We assume the presence of at least one variable in IBP system, otherwise we don't need functional reconstruction at all. By `real time', we refer to wall clock time, which might be shorter than the total CPU time for a multi-threaded process. $T_0$ can be determined from the information of sample points, the number of threads and the average sample time per probe per CPU. In line 5, setting the time limit for the search stage is essential in scenarios where searching for block-triangular relations might consume more time than the best available strategy. The parameter $F$ provides control over acceptable wasted time. The estimation in line 9 includes the real time for generating numerical IBP inputs, fitting the block-triangular form and computing probes using the block-triangular form. Line 11 updates the shortest real time for computing probes based on all currently available block-triangular forms.

\section{Benchmarks}\label{sec:benchmarks}

The latest version of \bl ~can be downloaded from
\begin{center}
	\url{https://gitee.com/multiloop-pku/blade}
\end{center}
Users can refer to the instructions provided in README.md for installation and utilization of the package.

In this section, we present benchmarks of \bl.
The calculations are performed on cluster nodes equipped with Intel Xeon Gold 6226R processors.
Note that numbers should be read with uncertainties due to different qualities of CPUs and fluctuating performance depending on workload of the computer.

\subsection{Three-loop four-point diagram with one massive external line}

\begin{figure}
	\centering
	\includegraphics[width=0.5\linewidth]{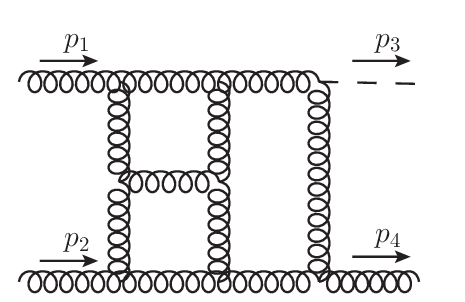}
	\caption{{\tt higgsjet}: a planar diagram that occurs in $\mathrm{N}^3$L$\mathrm{O}_{\mathrm{HTL}}$ QCD corrections to $
		gg\rightarrow Hg$}
	\label{fig:higgsjet_diagram}
\end{figure}

To illustrate the power of \bl, we start with the IBP reduction of the planar diagram shown in Fig.~\ref{fig:higgsjet_diagram}, which occurs in the $\mathrm{N}^3$LO QCD corrections to Higgs plus jet production in the heavy-top-limit and other similar processes.
The inverse propagators are:
\begin{equation}
	\begin{aligned}
		D_1 &= l_1^2, \quad D_2=(p_1+l_1)^2, \quad D_3 = (p_1+l_1+l_2)^2,
		\\
		D_4&=(p_1+p_2+l_1+l_2)^2, \quad
		D_5 =l_3^2, \\
		D_6 &=(-l_1-l_3)^2, \quad D_7=l_2^2,\quad D_8=(-l_1-l_2-l_3)^2,
		\\
		D_9&=(p_1+p_2-l_3)^2, \quad
		D_{10}=(p_3-l_3)^2,
		\\ D_{11}&=(p_2+l_2)^2,\quad D_{12}=(l_1+l_2)^2,\quad  D_{13}=(p_2+l_3)^2,
		\\
		D_{14}&= (p_3+l_1)^2,\quad D_{15}=(p_3+l_2)^2,
	\end{aligned}
\end{equation}
where the last five are ISPs.

The scalar products among independent external momenta are defined as:
\begin{align*}
	& p_1^2=p_2^2=0, \quad p_3^2=m_H^2, \quad
	(p_1+p_2)^2=s, \\
	& \quad (p_1-p_3)^2=t, \quad
	(p_2-p_3)^2=m_H^2-s-t.
\end{align*}
The family involves 121 MIs. The reduction coefficients depend on three variables $\epsilon,m_H^2$ and $t$, with $s$ set to 1.

\begin{table}[]
	\captionsetup{justification=raggedright,singlelinecheck=false}
	\caption{Acceleration of the block-triangular form in the {\tt higgsjet} family (Fig.~\ref{fig:higgsjet_diagram}). The $r_{max}$ represents the maximal rank of target top-sector integrals with no increased propagator powers. The block-triangular form can be closed with the option ``BladeMode" $\rightarrow$ None.}
	\label{tab:higgsjet_cpuh}
	\centering
	\begin{tabularx}{\linewidth}{|c|*{2}{>{\centering\arraybackslash}X|}}
		\hline
		$r_{max} $ & \makecell[c]{ CPU $\cdot$ h (With\\ block-triangular form)} & \makecell[c]{ CPU $\cdot$ h (Without\\ block-triangular form)} \\ \hline
		3 & 60 & 1800  \\ \hline
		4 & 180 & 11000
		\\ \hline
	\end{tabularx}
\end{table}

In Tab.~\ref{tab:higgsjet_cpuh}, we list the CPU core hours required for IBP reduction of top-sector integrals using the function ``BLReduce'' within \bl.
The parameter $r_{max}$, representing the largest rank of top-sector target FIs, is used to control the reduction's complexity.
Notably, in the considered example, the block-triangular form enhances the IBP reduction efficiency by 1-2 orders of magnitude.
Furthermore, the block-triangular form's advantages become more pronounced as the reduction task grows in complexity.

\begin{table}[]
	\captionsetup{justification=raggedright,singlelinecheck=false}
	\caption{Detailed account of time consumed in computing probes in the {\tt higgsjet} family (Fig.~\ref{fig:higgsjet_diagram}) with $r_{max}=3$. $t_{\mathrm{IBP}}$ and $t_{\mathrm{BL}}$ represent time required to compute a single numerical probe using plain IBP or the block-triangular form. $N_{\mathrm{search}}$ corresponds to the number of samples required to search the block-triangular form. $N_{\mathrm{fit}}$ indicates the number of samples necessary for subsequent construction of block-triangular form over a finite field. $N_{\mathrm{recon}}$ represents the number of samples required for functional reconstruction over a finite field. $N_{\mathrm{primes}}$ signifies the number of finite fields needed to reconstruct
		the rational numbers.}
	\label{tab:higgsjet_compute_probes}
	\centering
	\begin{tabularx}{\linewidth}{|*{6}{>{\centering\arraybackslash}X|}}
		\hline
		$t_{\mathrm{IBP}}$ & $t_{\mathrm{BL}}$ & $N_{\mathrm{search}}$& $N_{\mathrm{fit}}$ & $N_{\mathrm{recon}}$ & $N_{\mathrm{primes}}$\\ \hline
		41 s & 0.1 s & 400 &150  & 42000 & 4 \\ \hline
	\end{tabularx}
\end{table}

~\newline

In Tab.~\ref{tab:higgsjet_compute_probes}, we use  $r_{max}=3$ as an example to discuss the time-consumption for computing probes in detail.
\bl ~utilizes 400 numerical IBP probes as input to search the block-triangular form in the first prime field. When proceeding to other primes, we leverage the knowledge gained from the search phase, thereby requiring only 150 numerical probes in a finite field.
Employing the block-triangular form, we compute a large number of numerical probes(42000) for functional reconstruction within a finite field and we use four primes to complete the rational reconstruction(with one prime used for validation).
The numerical IBP costs 41 second per probe whereas the block-triangular form is about 400 times faster than the numerical IBP(0.1s).
Hence, compared to numerical IBP, the block-triangular form accelerates probe computation by a factor of $(4*42000*41)/(4*42000*0.1 + (1*400+3*150)*41) \approx 133$.
This represents the best runtime improvement achievable by the block-triangular form when the CPU time for probes using block-triangular form is dominant.

\begin{table}[]
	\captionsetup{justification=raggedright,singlelinecheck=false}
	\caption{Detailed account of time consumed in reducing top-sector integrals in the {\tt higgsjet} family (Fig.~\ref{fig:higgsjet_diagram}) with $r_{max}=3$.}
	\label{tab:higgsjet_detail_time}
	\centering
	\begin{tabularx}{\linewidth}{|*{6}{>{\centering\arraybackslash}X|}}
		\hline
		$N_{\mathrm{threads}}$  &  $T_{\mathrm{genIBP}}$ &  $T_{\mathrm{genDeg}}$ & $T_{\mathrm{search}}$ & $T_{\mathrm{fit}}$ per prime &  $T_{\mathrm{tot}}$  \\ \hline
		4 & 4h 20min & 1h 25min& 3h 30min &300s  & 14h 45min \\ \hline
		8 & 3h 50min & 1h 25min & 3h & 170s &10h 50min \\ \hline
	\end{tabularx}
\end{table}

~\newline

\begin{table}[]
	\captionsetup{justification=raggedright,singlelinecheck=false}
	\caption{The impact of the ``FilterLevel'' option in the {\tt higgsjet} family (Fig.~\ref{fig:higgsjet_diagram}) with $r_{max}=3$. In the first column, ``r'' denotes rank, ``d'' denotes dots, ``FL'' denotes ``FilterLevel''. The ``Mem'' in the third column represents the memory used to load and trim the raw IBP system, usually the maximum memory consumption in the calculation.}
	\label{tab:higgsjet_filterlevel}
	\centering
	\begin{tabularx}{\linewidth}{|c|*{5}{>{\centering\arraybackslash}X|}}
		\hline
		Seeds & \makecell[c]{\#Eqs.}  & Mem & \makecell[c]{\#Eqs.\\(trim)}&  $t_{\mathrm{IBP}}$ \\ \hline
		\makecell[c]{r: 3, d: 1,\\FL: $3$ (default) } & 2.9 M & 25 GB & 1.2 M &41 s 
		\\ \hline
		\makecell[c]{r: 3, d: 1,\\FL: $\infty$}  & 8.1 M & 59 GB & 2.3 M &60 s 
		\\ \hline
		\makecell[c]{r: 3, d: 1,\\FL: $\{1,0,-1,-2\}$}  & 0.8 M & 9 GB & 0.4 M & 25 s 
		\\ \hline
		\makecell[c]{r: 3, d: 0,\\FL: $\{1,1,0,-1,-2\}$}  & 0.28 M & 3 GB & 0.20 M &12 s 
		\\ \hline
	\end{tabularx}
\end{table}

\begin{table}[]
	\captionsetup{justification=raggedright,singlelinecheck=false}
	\caption{The impact of the ``SpanningReduce" option in the {\tt higgsjet} family (Fig.~\ref{fig:higgsjet_diagram}).}
	\label{tab:higgsjet_spanning}
	\centering
	\begin{tabularx}{\linewidth}{|c|*{5}{>{\centering\arraybackslash}X|}}
		\hline
		$r_{max}$  &  CPU $\cdot$ h &
		Mem \\ \hline
		3  & 35 &  7 GB
		\\ \hline
		4  & 110 &  21 GB
		\\ \hline
	\end{tabularx}
\end{table}

Due to the fast evaluation of the block-triangular form, the cost of computing probes only plays a minor role in the \bl.
Tab.~\ref{tab:higgsjet_detail_time}, presents the time distribution among various components in \bl.
The generation and trimming of the IBP system takes approximately 4 hours and 20 minutes.
In this example, the maximal dots of seed integrals should be greater than the dots of target integrals by one, namely, $d=1$.
The advantage of multivariate reconstruction algorithm employed in {\tt FiniteFlow} lies in its capability to generate necessary samples using the degree information obtained through univariate  reconstruction, prior to any multivariate reconstruction process.
Consequently, sample computation can be fully parallelized.
Generation of degree information accouts for 1 hour and 25 minutes, constituting 10\% of the total runtime.
\bl ~adopts adaptive-search strategy to search for block-triangular form and the optimal choice turns out to be the fully analytic block-triangular form.
$T_{\mathrm{search}}$ accounts for the total time of the adaptive-search process, including the time for block construction, computing numerical IBP samples as input and solving the linear system(Eq.~\eqref{eq:search_constraints}).
Its relevance is confined to the first prime field due to unknown explicit structure of block-triangular form.
The time required for solving the linear system to obtain the block-triangular form, denoted as $T_{\mathrm{fit}}$, in other primes, is considerably shorter.

The CPU time is contingent upon the number of threads used in computation, owning to the CPU inefficiency exhibited in other processes in \bl.
For instance, as depicted in Tab.~\ref{tab:higgsjet_detail_time}, utilizing 8 cores results in a 30\% reduction in runtime in this task.
The reason behind is that the time required for generating symmetry relations is longer compared to IBP and LI identities, particularly noticeable in seed integrals with high numerator powers.
Consequently, the CPU efficiency diminish as some cores remain idle, awaiting the computation of the most intricate symmetry relations.
This is also the case in the search process where some cores await the search for the most complex block-triangular relations.
Additionally, the trimming stage of {\tt FiniteFlow} only utilize one core.
To reduce the CPU time needed for complete reduction of a physical problem (often involving numerous integral families), it is advisable to allocate an appropriate number of threads for a single reduction job, or processes before computing probes.
A good choice is to ensure that the run time for probes is comparable with that of other processes.
The global variable ``BLNthreads'' determines the number of threads and
\bl ~would access the available number of threads (not exceeding BLNthreads) by computing the additional memory usage upon increasing a single thread, given that the main memory usage of numerical IBP is usually very expensive.
It's important to highlight that the block-triangular form usually optimally engages all threads in the reduction job due to its compact size.

The impact of the ``FilterLevel'' option is outlined in Tab.~\ref{tab:higgsjet_filterlevel}.
With the default option ``FilterLevel'' $\rightarrow $ 3, the sample time for numerical IBP decreases by approximately 30\%, compared to that of ``FilterLevel'' $\rightarrow \infty$.
Moreover, it necessitates less memory for numerical IBP.
The improvement can be attributed to the reduced size of IBP system.
The ``FilterLevel'' is set to 3 (equivalently, $\{0,0,0,-1\}$), signifying that the maximal rank of seed integrals satisfying $t \in \{10,9,8\}$ is 3 while those satisfying $t\leq 7$ possess a maximum rank of 2.
Setting ``FilterLevel" to infinity designates a maximum rank of 3 for seed integrals across all sectors.
The default setting requires fewer seed integrals, adequate for reducing top-sector integrals with rank no greater than 3, namely, $r\leq 3$.
Consequently, the size of the original IBP system decreases from $8.1*10^6$ to $2.9 *10^6$ and the size of trimmed IBP system decreases from $2.3*10^6$ to $1.2*10^6$.
By tuning the seeds more boldly, specifically $\{1,1,0,-1,-2\}$, we can further diminish memory usage by a factor of approximately 8.3 and decrease $t_{\mathrm{IBP}}$ by approximately 3.4, compared to those with the default set of ``FilterLevel''. Additionally, the seeding strategies used at $r_{max}=3$ can effectively extend to $r_{max} = 6$, which is crucial for Higgs plus jet amplitudes.
Empirically, ``FilterLevel'' performs well for reducing target integrals generated by the operator extension, especially in problems involving three loops or higher.

The effect of the ``SpanningReduce" option is shown in Tab.~\ref{tab:higgsjet_spanning}. \bl ~generates 22 spanning sectors, with 6 having 4 propagators at the lowest level sector, 15 containing 5 propagators, and 1 spanning sector containing 6 propagators. 
Choosing the default set for ``FilterLevel'', we find the CPU core hours are comparable to those without ``SpanningReduce" in Tab.~\ref{tab:higgsjet_filterlevel}, but the memory usage is reduced by a factor of 3 to 4.

When compared to other reduction packages available on the market, such as {\tt FiniteFlow}\cite{Peraro:2019svx} + {\tt LiteRed}\cite{Lee:2013mka}, {\tt Fire}\cite{Smirnov:2023yhb}, and {\tt Kira}\cite{Klappert:2020nbg}, we observed that the computational time of \bl, when not utilizing the block-triangular form, is similar to those packages. For example, when compared to {\tt Kira2.2}\cite{Klappert:2020nbg} + {\tt Firefly2.0}\cite{Klappert:2020aqs}, \bl ~typically computes a single sample faster but requires more numerical samples for functional reconstruction, resulting in overall effects that are comparable. This trend is consistent across other examples, leading us to focus on benchmarks specifically within \bl.

\subsection{Four-loop two-point diagram with one massive internal particle}

\begin{figure}
	\centering
	\includegraphics[width=0.3\textwidth]{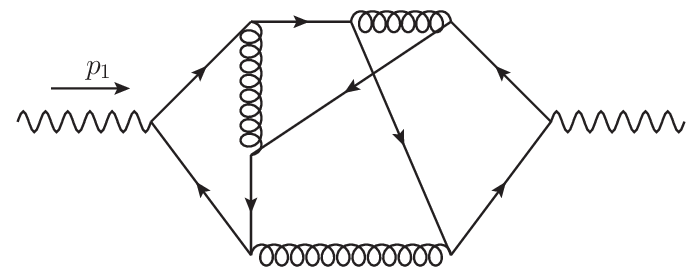}
	\caption{{\tt fsf3}: a non-planar diagram of forward scattering.}
	\label{fig:fsf3}
\end{figure}

Our next benchmark is an example from study of $\mathrm{N}^3$LO QCD corrections to heavy-quark pair production at lepton colliders, shown in Fig.~\ref{fig:fsf3}.
The inverse propagators for {\tt fsf3} are:
\begin{equation}
	\begin{aligned}
		D_1 &= l_1^2, \quad D_2=l_2^2, \quad D_3=l_3^2, \quad D_4=l_4^2-m_t^2,
		\\
		D_5&=(p_1+l_4)^2-m_t^2, \quad
		D_6=(l_1+l_4)^2-m_t^2,
		\\ D_7&=(p_1+l_2+l_4)^2-m_t^2,
		\quad D_8=(l_1+l_2+l_4)^2-m_t^2,
		\\
		D_9&=(p_1+l_2+l_3+l_4)^2-m_t^2,
		\\
		D_{10}&= (l_1+l_2+l_3+l_4)^2-m_t^2,
		\\
		D_{11}&=(p_1+l_1+l_2+l_3+l_4)^2-m_t^2,
		\\
		D_{12}&=(p_1+l_1)^2, \quad
		D_{13}=(p_1+l_2)^2,\quad D_{14} = (l_2+l_3)^2,
	\end{aligned}
\end{equation}
where the last three are ISPs.

The scalar product of external momenta is defined as:
\begin{equation}
	p_1^2 = s.
\end{equation}
The family involves 369 MIs. The reduction coefficients dependent on two variables $\epsilon$ and $m_t^2$ with $s$ set to 1.

As depicted in Tab.~\ref{tab:fsf3_cpuh}, significant enhancement of IBP reduction using the block-triangular form has been observed in this example. A detailed account of time-consumption involved in computing probes and complete reduction with $r_{max}$=4 is shown in Tab.~\ref{tab:fsf3_compute_probes} and Tab.~\ref{tab:fsf3_detail_time_blade} respectively.
Notably, the block-triangular form is 400 times faster than the numerical IBP, leading to a 69-fold reduction in the time required for probe computation.
The time required for generating degree information takes a very large proportion of the total time due to the low efficiency of numerical IBP.
To mitigate this, we can opt for a  specific setting to reduce the time overhead.
By utilizing the option ``BladeMode''$\rightarrow$ Full, \bl~will directly search the full-analytic block-triangular form instead of employing the adaptive search strategy. Consequently, the step of generating degree information using numerical IBP is skipped.
The degree information will be derived by solving the block-triangular form once it is successfully constructed.
This improvement is based on the observation that the block-triangular form with three or fewer parameters is very likely to be constructed.
In this example, $T_{\mathrm{genDeg}}$ is reduced to 1 hour.

\begin{table}[]
	\captionsetup{justification=raggedright,singlelinecheck=false}
	\caption{Acceleration of the block-triangular form in the {\tt fsf3} family (Fig.~\ref{fig:fsf3})}
	\label{tab:fsf3_cpuh}
	\centering
	\begin{tabularx}{\linewidth}{|c|*{2}{>{\centering\arraybackslash}X|}}
		\hline
		$r_{max} $ & \makecell[c]{ CPU $\cdot$ h (With\\ block-triangular form)} & \makecell[c]{ CPU $\cdot$ h (Without\\ block-triangular form)} \\ \hline
		3 & 120 & 1200  \\ \hline
		4 & 280 & 8000
		\\ \hline
	\end{tabularx}
\end{table}

\begin{table}[]
	\captionsetup{justification=raggedright,singlelinecheck=false}
	\caption{Time consumed in computing probes in the {\tt fsf3} family (Fig.~\ref{fig:fsf3}) with $r_{max}=4$.}
	\label{tab:fsf3_compute_probes}
	\centering
	\begin{tabularx}{\linewidth}{|*{6}{>{\centering\arraybackslash}X|}}
		\hline
		$t_{\mathrm{IBP}}$ & $t_{\mathrm{BL}}$ & $N_{\mathrm{search}}$&
		$N_{\mathrm{fit}}$& $N_{\mathrm{recon}}$ & $N_{\mathrm{primes}}$ \\ \hline
		440 s & 1.1 s& 128 & 64 & 6041 & 8\\ \hline
	\end{tabularx}
\end{table}

\begin{table}[]
	\captionsetup{justification=raggedright,singlelinecheck=false}
	\caption{Time consumed in reducing top-sector integrals in the {\tt fsf3} family (Fig.~\ref{fig:fsf3}) with $r_{max}=4$.}
	\label{tab:fsf3_detail_time_blade}
	\centering
	\begin{tabularx}{\linewidth}{|*{6}{>{\centering\arraybackslash}X|}}
		\hline
		$N_{\mathrm{threads}}$& $T_{\mathrm{genIBP}}$ &  $T_{\mathrm{genDeg}}$ & $T_{\mathrm{search}}$ & $T_{\mathrm{fit}}$ per prime &  $T_{\mathrm{tot}}$ \\ \hline
		4 & 13 h & 26 h& 8 h &300 s & 70 h \\ \hline
	\end{tabularx}
\end{table}

\subsection{Two-loop four-point diagram with two massive internal particle}\label{sec:twoloop_fourpoint}

\begin{figure}
	\centering
	\includegraphics[width=0.6\linewidth]{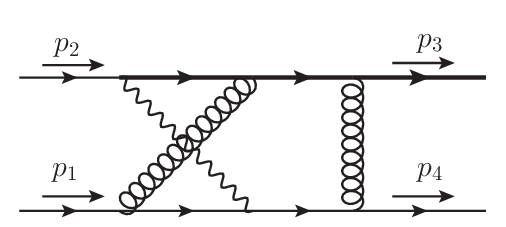}
	\caption{{\tt topo5}: A non-planar diagram that occurs in single top production. Wavy line, spiralled, bold line and line represent W boson, gluons, top-quark and light quarks.}
	\label{fig:topo5_diagram}
\end{figure}

We also study the IBP reduction of the non-planar double box {\tt topo5} which occurs, e.g. in the $\mathrm{N}^2\mathrm{LO}$ corrections to single top production. The set of propagators is chosen as
\begin{equation}
	\begin{aligned}
		D_1 &= l_1^2, \quad D_2 = l_2^2, \quad D_3 = (p_4-l_1)^2,\quad D_4=(p_1-l_2)^2,
		\\
		D_5 &=(p_3+l_1)^2-m_t^2,\quad D_6= (p_3+l_1-l_2)^2 -m_t^2,
		\\
		D_7 &= (-p_2+p_3+l_1-l_2)^2-m_W^2,
		\\
		D_8 &= (l_1-p_1)^2,\quad D_9 = (l_2-p_4-p_2)^2,
	\end{aligned}
\end{equation}
where the last two are ISPs.

The scalar products among external momenta are defined as:
\begin{equation}
	\begin{aligned}
		& p_1^2 = 0,\quad p_2^2=0, \quad p_4^2 = 0, \quad
		(p_1+p_2)^2 = s,\\
		& (p_4-p_1)^2 =t,\quad (p_4-p_2)^2 = (-s-t+m_t^2)
	\end{aligned}
\end{equation}
The family involves 76 MIs. The reduction coefficients depend on four parameters, $\epsilon , s, t$ and $m_W^2$, with $m_t$ set to 1.

We conduct the reduction of top-sector integrals with $r_{max}=4$.
The full-analytic block-triangular form is too hard to search with the default options and \bl ~utilizes the semi-analytic block-triangular form with respect to $\{t, \epsilon, m_W^2\}$ to compute probes for reconstruction.
Specifically, under one prime field, \bl ~constructs approximately 150 sets of block-triangular forms, each assigned a distinct numerical value for $s$. It solves these relations by substituting various numerical values for $\{t, \epsilon, m_W^2\}$. These numerical samples are properly used for reconstruction.
As shown in Tab.~\ref{tab:topo5_detail_time}, \bl ~utilizes around 115,000 numerical IBP probes to construct the block-triangular form and computes about 8.9 million probes across five primes to complete the reduction.
The block-triangular form significantly enhances probe computation efficiency, making the total computational time predominantly limited by multivariate reconstruction, which accounts for approximately 80\% of the total CPU core hours (290).

Given the complexity of reconstruction, it is suggestive to set numeric values to mass scales in \bl, which is sufficient to give high precision predictions in phenomenological applications.
In this example, we set $\frac{m_W^2}{m_t^2} = \frac{14}{65}$.
As illustrated in Tab.~\ref{tab:topo5_detail_time}, this assignment results in a reduction by a factor of 22 in the number of numerical samples required for reconstruction, as well as in the  total CPU core hours. Consequently, the time required for reconstruction becomes negligible (3\%).
It is also observed that assigning mass values effectively reduces the $t_{\mathrm{IBP}}$ and $t_{\mathrm{BL}}$, owning to the simplification of linear relations.

The majority of the computational time, around 60\%, is consumed by the probe computation and subsequent reconstruction. Yet, what accounts for the remaining 40\%?
The reasons are twofold: Firstly,  the heavy disk read/write tasks, induced by a substantial amount of data, result in increased processing time. Secondly, the time required for fitting block-triangular form ($T_{\mathrm{fit}}$) becomes more substantial as the processing time of other components decreases. Additionally, the CPU efficiency during the fitting process is comparatively low.
These factors persist, but become notably pronounced, particularly when employing the semi-analytic block-triangular form. We leave the improvements for future work.

\begin{table}[]
	\captionsetup{justification=raggedright,singlelinecheck=false}
	\caption{Time consumed in reducing top-sector integrals in the {\tt topo5} family (Fig.~\ref{fig:topo5_diagram}) with $r_{max}=4$.}
	\label{tab:topo5_detail_time}
	\centering
	\begin{tabularx}{\linewidth}{|c|c|c|*{4}{>{\centering\arraybackslash}X|}}
		\hline
		Mode  & $t_{\mathrm{IBP}}$ & $t_{\mathrm{BL}}$ & $N_{\mathrm{fit}}$ & $N_{\mathrm{recon}}$ & $N_{\mathrm{primes}}$ & CPU$\cdot$ h\\ \hline
		default & 0.48 s & 0.03 s & 23000 & 1788811 & 5 & 290 \\ \hline
		$\frac{m_W^2}{m_t^2} = \frac{14}{65}$ & 0.32 s & 0.012 s& 6500 &81649 & 8 & 13 \\
		\hline
	\end{tabularx}
\end{table}

\subsection{Two-loop five-point diagram with one massive external line}

\begin{figure}
	\centering
	\includegraphics[width=0.7\linewidth]{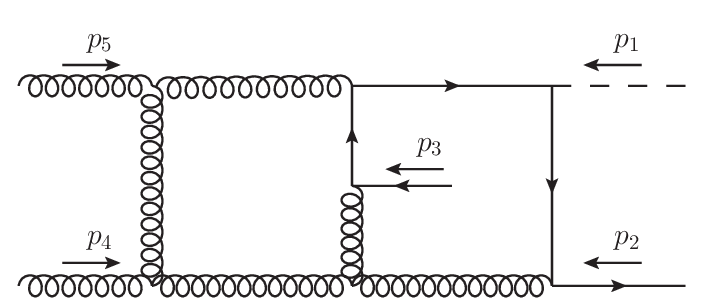}
	\caption{{\tt dpmass}: a non-planar diagram that occurs in $Hb\bar{b}$ production.}
	\label{fig:Hbbar_diagram}
\end{figure}

Our last benchmark comes from the study of two loop five point scattering amplitudes with one massive external line, which has garnered significant attention in recent years \cite{Badger:2021nhg,Badger:2021ega,Badger:2022ncb,Abreu:2021asb,Hartanto:2022qhh,Abreu:2023rco}. Despite the successful resolution of all master integrals \cite{Abreu:2023rco}, integral reduction continues to pose a challenge.
We select the non-plannar diagram, shown in Fig. \ref{fig:Hbbar_diagram} to demonstrate the potential application and efficacy of \bl.
The set of inverse propagators is chosen as
\begin{equation}
	\begin{aligned}
		D_1 &= l_1^2, \quad D_2 = (l_1+p_1)^2, \quad D_3 = (l_1+p_1+p_2)^2,
		\\
		D_4 &=l_2^2, \quad
		D_5 =(l_2-p_1-p_2-p_3-p_4)^2,
		\\
		D_6&= (l_2-p_1-p_2-p_3)^2, \quad
		D_7= (l_1+l_2-p_3)^2,
		\\
		D_8 &= (l_1+l_2)^2, \quad D_9 = (l_1+p_4)^2,
		\\
		D_{10} &= (l_2+p_2)^2, \quad D_{11} = (l_2+p_1)^2,
	\end{aligned}
\end{equation}
where the last three are ISPs.

The five momenta $p_i$ are subject to on-shell and momentum conservation conditions,
\begin{align}
	& p_1^2 = m^2,\quad p_i^2=0, ~i=2,...,5  \\ \nonumber
	& \sum_{i=1}^5 p_i = 0.
\end{align}
They give rise to five independent invariants \{$s_{12}$,$s_{23}$,$s_{34}$, \\ $s_{45}$,$s_{15}$\} with $s_{ij}$=$(p_i+p_j)^2$, and we set $s_{12}$=1. The family involves 142 MIs.

\begin{table}[]
	\captionsetup{justification=raggedright,singlelinecheck=false}
	\caption{Time consumed in computing probes in the  {\tt dpmass} family (Fig.~\ref{fig:Hbbar_diagram}) with $r_{max}=5$.}
	\label{tab:dpmass_detail_time}
	\centering
	\begin{tabularx}{\linewidth}{|c|*{7}{>{\centering\arraybackslash}X|}}
		\hline
		$r_{max}$  & $t_{\mathrm{IBP}}$ & $t_{\mathrm{BL}}$ & $N_{\mathrm{search}}$ &$N_{\mathrm{fit}}$ & $N_{\mathrm{recon}}$ \\ \hline
		5 & 6 s & 0.16 s &2000 & 1000  & ?$10^5$  \\ \hline
	\end{tabularx}
\end{table}

A key technique in our benchmark is setting $\epsilon$ to numerical values, for instance, $\epsilon = 1/1000$, as the $\epsilon$ dependence can be reconstructed at the final stage of a computation such as the cross section \cite{Liu:2022mfb,Liu:2022chg}.
It is crucial to utilize the mass dimension constraints of linear relations among Feynman integrals. This involves making polynomial ansatz similar to Eq.~\eqref{eq:ansatz-of-rel}. Specifically, we employ the following options:
``VariableGroup'' $\rightarrow$ \{\{$s_{45}, s_{34}, s_{23}, s_{15}, m^2$\}\}, ``IntegralWeight'' $\rightarrow$ \{``dimension"\},  ``VariableWeight'' $\rightarrow$ \{``uniform"\}.
This ansatz mode significantly reduces the number of unknowns in the search algorithm, making it feasible to search for the block-triangular form.

For target top-sector FIs with $r_{max}=5$, it took approximately 5 days to search for the block-triangular form in the first prime-field using 4 cores. However, fitting the block-triangular form when proceeding to other prime fields or $\epsilon$ value only required 8 hours. This time could be further reduced to 1.5 hours if we use 2000 numeric IBP as input, namely, $N_{\mathrm{fit}}=2000$, because more numeric IBP give more constraints for each master integral, thus making the linear equations generated from a single master integral as sparse as possible after backward substitution (see Alg.~\ref{alg:search_algorithm}). As depicted in Tab.~\ref{tab:dpmass_detail_time}, the obtained block-triangular form is about 38 times faster than the numeric IBP.
Probes for reconstruction are indeterminate due to exhausting 1.5 TB memory during sample point generation. Improvements in reducing numerical probes and optimizing multivariate reconstruction would be beneficial. Alternatively, solving the rational reconstructed block-triangular form in floating-point  numbers can be a favorable option given the manageable precision loss and high precision numerical result of master integrals provided by {\tt AMFlow} \cite{Liu:2017jxz,Liu:2020kpc,Liu:2021wks,Liu:2022mfb,Liu:2022chg}.

\section{Summary and Outlook}\label{sec:summary}

In this article, the fully automated Feynman integral reduction package \bl ~is presented together with various explicit examples. Armed with the method of block-triangular form \cite{Liu:2018dmc,Guan:2019bcx}, \bl ~typically enhances the IBP reduction efficiency by 1-2 orders of magnitude.
Furthermore, \bl ~has many distinctive features, which make it applicable in many general cases.

In this version of \bl,~the computational time is no longer exclusively limited by computing probes thanks to the high efficiency of the block-triangular form.
Other procedures, including generating IBP equations, deriving degree information, searching for/fitting the block-triangular form, and performing multivariate reconstruction, also hold pivotal significance, each contributing a proportion tailored to the specific problems encountered. Therefore, comprehensive optimization is necessary for further improvements.

We aim to provide an interface supporting user-defined IBP systems, potentially outperforming the standard IBP system generated by \bl, in terms of memory usage or numerical efficiency.
For instance, a system generated utilizing syzygy equations is highly desired. The natural next step would be to incorporate the Singualr/Mathematica based package {\tt NeatIBP} \cite{Wu:2023upw} into \bl. 

In future, we plan to migrate the existing Mathematica-based package to an open-source language implementation. The performance is expected to be improved as well, such as generating IBP equations and I/O operations. Utilizing univariate block-triangular form for deriving degree information would be advantageous.
We are continuously exploring improved schemes for integral extension and polynomial ansatz. These efforts aim to enhance the efficiency of searching for the block-triangular form and provide deeper insights into the properties of Feynman integrals.
Recycling the numerical IBP probes generated in the intermediate step of the adaptive search stage in subsequent searches procedure will further reduce the number of required IBP probes. We are also exploring the potential to combine the partial-fractioned reconstruction technique \cite{Chawdhry:2023yyx} with \bl.

\section*{Acknowledgements}
We thank Li-Hong Huang, Xiang Li, Bernhard Mistlberger, Tiziano Peraro, Ao Tan, Zi-Hao Wu, and Yang Zhang for many useful communications and discussions. We thank Huan-Yu Bi, Long Chen, Xiang Chen, Yuan-Hong Guo, Chuan-Qi He, Rui-Jun Huang, Xu-Hang Jiang, Xiang Li, Huai-Min Yu and Yu-Dong Zhang for providing numerous invaluable feedback on the package.
The work was supported in part by the National Natural Science Foundation of
China (Grants No. 12325503, No. 11975029), the National Key Research and Development Program of China under
Contracts No. 2020YFA0406400, and the High-performance Computing Platform of Peking University. The research of XG was also supported by the United States Department of Energy, Contract DE-AC02-76SF00515. The research of XL was also supported by the ERC Starting Grant 804394 \textsc{HipQCD} and by the UK Science and Technology Facilities Council (STFC) under grant ST/T000864/1.
{\tt JaxoDraw}~\cite{BINOSI200476} was used to generate Feynman diagrams.

\appendix

\section{Propagators}

Feynman integrals in \bl ~are considered in Minkowski space, $l^2$, of the denominators instead of `Euclidean' space, $-l^2$.

We associate to each denominator a number `pre' to indicate Feynman prescription,
\begin{align*}
	\begin{cases}
		D_j(l_i)+\mi 0, & \textrm{if} ~ pre=1,\\
		D_j(l_i), & \textrm{if} ~ pre=0, \\
		D_j(l_i)-\mi 0, &  \textrm{if} ~ pre=-1.
	\end{cases}
\end{align*}
`pre' is set to 1 by default.
Symmetry is detected among denominators that have identical prescriptions.
More details can be found in the ``example/1\_prescription'' folder.

Propagators that can be written as $q^2 - c$ or $q\cdot p - c $ is called standard propagators in \bl.
Here $q$ is a linear combination of loop momenta and external momenta, and $p$ is a linear combination of external momenta.
The current version of \bl ~will not detect symmetry relations if other general propagators are present.

\section{Discretized Symmetries}
\label{app}

\subsection{Point group of Feynman integral}

\begin{figure}
	\centering
	\includegraphics[width=\linewidth]{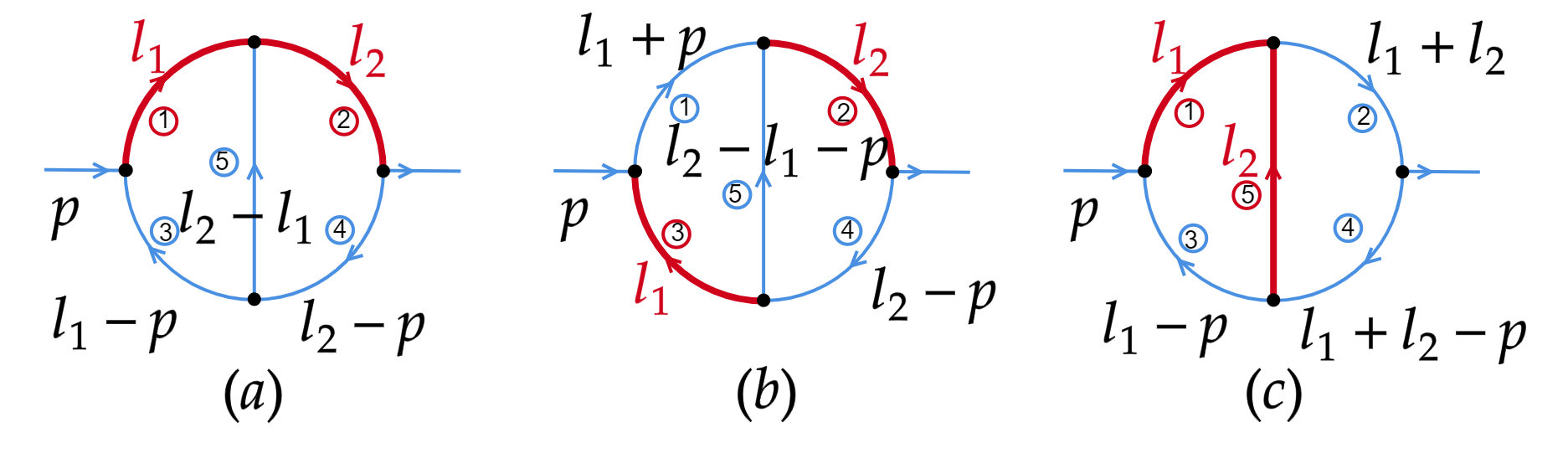}
	\caption{The choices of spanning tree (blue thin lines) and loop base (red thick lines) in a 2-loop self-energy diagram}
	\label{fig:st}
\end{figure}

Let us consider a Feynman integral with $P$ propagator denominators, where each denominator depends on a propagating momentum $q_a$, defined as
\begin{equation} \label{eq:propmom}
	q_a = \sum_{i=1}^L \alpha_a^i l_i + \sum_{j=1}^E \beta_a^j p_j, \quad a=1,\ldots,P.
\end{equation}
A {\it branch}  is defined by a set of propagators that share the same loop momenta information. For instance, in Fig.~\ref{fig:st}(a), there are three branches defined by the propagators \{\circled{1},\circled{3}\},  \{\circled{2},\circled{4}\}, and \{\circled{5}\}, respectively.

In Section \ref{sec:finitetransf}, we highlighted that linear transformations of loop momenta enable us to relate any given Feynman integral to its canonical form, which we will define next. The only condition we have imposed so far on the canonical form is that each loop momentum $l_i$ needs to serve as the propagating momentum of one of the propagators, as much as possible. We refer to the propagator with propagating momentum $l_i$ as the $i$-th {\it loop base}. However, there can be different choices of loop bases, and even for a given choice of loop bases, there can be different ways to label each propagating momentum.
Starting from a candidate for the canonical form, which means a choice of loop bases has already been made, we can generate many more candidates by applying the following operations:
\begin{itemize}
	\item {Reverse} propagating momentum of each propagator:
	\begin{equation}
		q_a \to q_a' = -q_a,
	\end{equation}
	which is just a matter of defining the orientation of the momentum.
	This results in $2^P$ choices.
	
	\item {Reverse} each loop momentum:
	\begin{equation}
		l_i \to l_i' = -l_i ,
	\end{equation}
	resulting in $2^L$ choices.

	\item Choose loop bases from different propagators. The number of choices equals the number of terms in the 1st Symanzik polynomial, denoted as \#$U$. For example, the diagram in Fig. \ref{fig:st} with internal lines labeled as shown has the 1st Symanzik polynomial
	\begin{equation}
		\begin{aligned}
			&U(x_1,x_2,x_3,x_4,x_5) \\
			=\,& (x_1+x_3)(x_2+x_4)
			+ (x_1+x_2+x_3+x_4)x_5
			\\ =\,& x_1 x_2 + x_1 x_4 + x_3 x_2 + x_3 x_4+ x_1 x_5 + x_2 x_5 \\
			& + x_3 x_5 + x_4 x_5,
		\end{aligned}
	\end{equation}
	which has 8 terms.
	In Fig.\ref{fig:st}, the choices of loop bases in diagrams (a), (b) and (c) correspond to the $1$st term $x_1 x_2$, 3rd term $x_3 x_2$, and 5th term $x_1 x_5$, respectively. Note that this counting is correct if we only choose double propagators as loop bases. 
	
	\item {Permutation} of loop momenta:
	\begin{equation}
		l_i \to l_i'=l_{\sigma(i)},
	\end{equation}
	where $\sigma$ is any permutation operation applying to the $L$ loop momenta. This results in $L!$ choices.
\end{itemize}
Thus, the total number of candidates for the canonical form is $2^P \times 2^L \times \# U \times L!$, which is usually very large, making it impractical to enumerate each choice and then sort them. Fortunately, since these operations form a point group with a well-organized structure, we can find a representative candidate with much less effort. 

\subsection{Canonical form of Feynman integral}

We assume the Feynman integral under consideration is already a candidate for the canonical form. Otherwise, we can simply choose any loop bases by selecting a term in the 1st Symanzik polynomial and transform the integral to a candidate form. Our algorithm to find the canonical form, which is a representative candidate among all possibilities, can be summarized in the following five steps:
\begin{description}
	\item[1)] \textbf{Orientations of propagating momenta.}
	
	For each propagator $q_a$ defined in Eq.~\eqref{eq:propmom}, we use a vector $v_a$ to denote it as $$v_a=(\alpha_a^1, \cdots, \alpha_a^L; \beta_a^1,\cdots,\beta_a^E) .$$
	We sort the list $\{v_a, -v_a\}$ in alphabetical order for each propagator and replace the vector $v_a$ by the last one of the sorting result. Then, if two propagators belong to the same branch, the first $L$ terms of their vectors must be the same, which enables us to classify propagators into branches.
	
	\begin{figure}
		\centering
		\includegraphics[width=\linewidth]{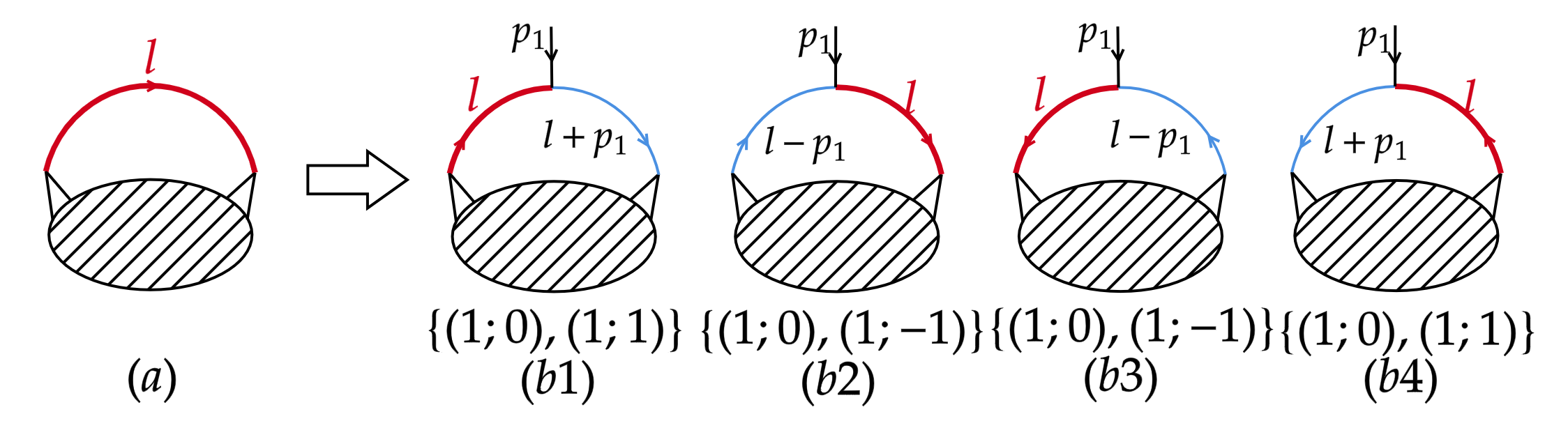}
		\caption{Partial loop bases in a length-$2$ branch}
		\label{fig:branch2}
	\end{figure}
	
	\item[2)] \textbf{Representative propagator in each branch.}
	
	For each branch, find the most special propagator by sorting all of them with different momenta orientations. The special one will be the candidate for the loop base in this branch. Loop momenta are irrelevant at this step.
	
	Usually, there is only one special propagator. However, for a length-2 branch with the same mass in both propagators, as shown in Fig. \ref{fig:branch2}, two pairs of representations are identical (b1, b4 and b2, b3). So, in the sub-diagram, neither is more special, and we need to retain both propagators for later consideration.
	
	\item[3)] \textbf{Loop bases.}

	Use the 1st Symanzik polynomial of the corresponding vacuum diagram, obtained by removing external legs, to determine all possible choices of loop bases. Sort the obtained expressions of the integrand for all possible choices to find a set of special choices. Note that we will not specify the label and orientation of loop momenta in this step, so only the total number of loop momenta and the absolute values of external momenta are relevant in the sorting procedure. For example, for an integral with three loops and two external momenta, the propagating momenta will be mapped to the following form during the sorting procedure:
	\begin{equation}
		\begin{split}
			q_1&=l_1-l_2+l_3+p_1 \to (3;1,0), \\ q_2&=l_2-l_1-p_2 \to (2;0,1).
		\end{split}
	\end{equation}
	
	\item[4)] \textbf{Orientations of loop momenta.}
	
	Beginning from the special choices in the previous step, iterate over all orientations of loop momenta, and sort the obtained expressions to find a class of special choices. As we will not specify the labels of loop momenta in this step, propagating momenta will be mapped to the following form during the sorting procedure:
	\begin{equation}
		\begin{split}
			q_1&=l_1-l_2+l_3+p_1 \to (1,1,-1; 1,0), \\
			q_2&=l_2-l_1-p_2 \to (1,-1;0,-1),
		\end{split}
	\end{equation}
	where the numbers before the semicolon are in an orderless manner, and the numbers after the semicolon are ordered.
	
	\item[5)] \textbf{Permutations of loop momenta.}
	
	Beginning from the special choices in the previous step, permute all loop momenta and sort the obtained expressions to find a class of special choices. All information for loop momenta and external momenta is specified in this step. For example, we have the following mapping relation:
	\begin{equation}\
		\begin{split}
			q&=l_1-l_2+l_3+p_1 \to (1,-1,1;1,0),
			\\
			q&=l_2-l_1-p_2 \to (-1,1;0,-1).
		\end{split}
	\end{equation}
	
\end{description}
Finally, we obtain a canonical form for the considered integral, as well as several linear transformations of loop momenta that can transform the original form to the canonical form. The complexity is roughly of the order $O( L!)$, insensitive to the number of external momenta. Although this complexity can be further reduced by splitting the fifth step to more steps, it is usually not necessary for practice use.

\subsection{Simplified symmetry relations}

With the canonical form, we can easily check whether two sectors are the same after applying a certain linear transformation of loop momenta. Furthermore, as there can be several transformations that keep the canonical form unchanged, known as ``automorphic" maps, they can generate linear relations of Feynman integrals (FIs) within the same sector. These are symmetry relations that will be used in FI reduction.

For any symmetry relation, the propagator denominators are permuted by $\sigma$ while the ISP undergoes a general linear transformation $D_a \to D_a' = \sum_{b}A_{ab}D_{b}$. More explicitly, we have
\begin{equation}
	\left\{
	\begin{matrix}\displaystyle
		D_{1} &\to& D^{'}_1 &=& D_{\sigma(1)}, \\
		&&\ldots\ldots&& \\
		D_{k} &\to& D^{'}_k &=& D_{\sigma(k)}, \\
		D_{k+1} &\to& D^{'}_{k+1} &=& \displaystyle\sum_{j=1}^{N} A_{k+1,j}D_{j}, \\
		&&\ldots\ldots&& \\
		D_{N} &\to& D^{'}_N &=& \displaystyle\sum_{j=1}^{N} A_{N,j}D_{j}.
	\end{matrix}\right.
\end{equation} 
For an integral
\begin{equation} \label{eq:symm}
	\int \mathrm{d}\mu_L \frac{D_{k+1}^{-\nu_{k+1}}\ldots D_{N}^{-\nu_n}}{D_1^{\nu_1}\ldots D_k^{\nu_k}},
\end{equation}
we can generate a symmetry relation by replacing $D_a$ with $\sum_{b}A_{ab}D_{b}$ for all values of $a$. However, if the rank $R=-\sum_{a=k+1}^N \nu_a$ is not very small, the length of this relation can be extremely long, scaling as $O(N^R)$  for $N\sim 10$ in cutting-edge problems.

In fact, the above integral can be re-expressed by any of the following kinds of symmetry relations:
\begin{align}
	& \int \mathrm{d}\mu_L \frac{(D_{k+1}^{(l)})^{-\nu_{k+1}}\ldots (D_{N}^{(l)})^{-\nu_N}}{D_1^{\nu_1}\ldots D_k^{\nu_k}} \nonumber\\
	= & \int \mathrm{d}\mu_L \frac{(D_{k+1}^{(r)})^{-\nu_{k+1}}\ldots (D_{N}^{(r)})^{-\nu_N}}{D_{\sigma(1)}^{\nu_{\sigma(1)}}\ldots D_{\sigma(k)}^{\nu_{\sigma(k)}}},
\end{align}
where $(D_a^{(l)},D_a^{(r)})$ can be set to either $(D_a,\sum_{b} A_{ab}D_b)$ or $(\sum_{b} A^{-1}_{ab}D_b,D_a)$.
By choosing $(D_a^{(l)},D_a^{(r)})$ for each value of $a$ to minimize the number of terms of the equation, we can find the simplest symmetry relation related to the integral in Eq.~\eqref{eq:symm}. This will roughly reduce the complexity from $O(N^R)$ to its square root, $O(\sqrt{N^R})$.

\newpage

\providecommand{\href}[2]{#2}\begingroup\raggedright\endgroup

\end{document}